\documentclass[a4paper,10pt,twoside,preprint,aps ,nofootinbib]{cpc-hepnp}

\usepackage{graphicx}
\pdfoutput=1
\usepackage{epsfig}
\usepackage{amsmath}
\usepackage{amsfonts}
\usepackage{amssymb}
\usepackage{color}
\usepackage{dcolumn}
\usepackage{hyperref}
\usepackage{multirow}
\usepackage{CJK}
\usepackage{array}
\usepackage{siunitx}
\usepackage{booktabs}

\providecommand{\U}[1]{\protect\rule{.1in}{.1in}}

\newcommand{\f}{\begin{equation}}
\newcommand{\ff}{\end{equation}}
\newcommand{\fa}{\begin{eqnarray}}
\newcommand{\ffa}{\end{eqnarray}}

\begin{document}

\title{\boldmath Modified regular black holes with time delay and 1-loop quantum correction
\thanks{We are very grateful to Dr. Hong Guo and Prof. Xiang Li for
helpful discussions. This work is supported in part by the Natural
Science Foundation of China under Grant No.~11875053 and 12035016.}}

\author{Yi Ling  $^{1,2}$ \email{lingy@ihep.ac.cn}
\quad Meng-He Wu  $^{3,4}$ \email{mhwu@sues.edu.cn}}

\maketitle

\address{
$^1$ Institute of High Energy Physics, Chinese Academy of Sciences, Beijing, 100049, China \\
$^2$ School of Physics, University of Chinese Academy of Sciences, Beijing, 100049, China \\
$^3$ School of Mathematics, Physics and Statistics, Shanghai University of Engineering Science, Shanghai, 201620, China \\
$^4$Center of Application and Research of Computational Physics, Shanghai University of Engineering Science, Shanghai, 201620, China }
\maketitle

\begin{abstract}
We develop the regular black hole solutions recently proposed in
\cite{Ling:2021olm} by incorporating the 1-loop quantum correction
to the Newton potential as well as a time delay between an
observer at the regular center and that at infinity. We define the maximal time
delay between the center and the infinity by scanning
the mass of black holes such that the sub-Planckian feature of
Kretschmann scalar curvature is preserved  during the whole process of evaporation.  We also compare the distinct behavior of
Kretschmann curvature for black holes with an asymptotically
Minkowski core and those with an asymptotically de-Sitter core,
including Bardeen and Hayward black holes. We expect that this
sort of regular black holes may provide more information about the
construction of effective metric for Planck stars.

\end{abstract}

\section{Introduction}
Before a complete theory of quantum gravity could be established,
it is quite desirable to construct regular black hole solutions,
which are non-singular everywhere and characterized by a finite
Kretschmann scalar curvature, to understand the final stage of
star collapse and black hole
evaporation\cite{Adler:2001vs,Amelino-Camelia:2004uiy,Ashtekar:2005cj,Amelino-Camelia:2005zpp,Ling:2005bq,Ling:2005bp,Bonanno:2000ep,Rovelli:2014cta}.
In literature
\cite{Bardeen:1968,Hayward:2005gi,Frolov:2014jva,Bogojevic:1998ma,Mazur:2001fv,Dymnikova:2001fb,Nicolini:2005vd,Falls:2010he,Bambi:2013ufa,Xiang:2013sza,Culetu:2013fsa,Culetu:2014lca,Simpson:2019mud,Ghosh:2014pba},
such regular black holes can be constructed by taking the quantum
effects of gravity into account in a heuristical manner, and can
be classified into two categories based on their asymptotical
behavior at the central core which now is regular. One is the
regular black hole with asymptotically de-Sitter core, which
includes the well known Bardeen black hole\cite{Bardeen:1968},
Hayward black hole\cite{Hayward:2005gi} as well as Frolov black
hole\cite{Frolov:2014jva}; the other is the regular black hole
with asymptotically Minkowski core, which was originally proposed
in \cite{Xiang:2013sza} and then was developed in
\cite{Culetu:2013fsa,Culetu:2014lca,Ghosh:2014pba,Simpson:2019mud}
. Recently, in \cite{Ling:2021olm}  we proposed an exponentially
suppressing form of gravity potential which is mass dependent such
that Kretschmann scalar curvature is bounded above by the Planck
mass density, regardless of the mass of the black hole.
Furthermore, we established a one-to-one correspondence between
regular black holes with Minkowski core and those with de-Sitter
core. This correspondence provides us a scheme to construct new
regular black holes with dS core as well.

With the goal to construct effective metrics to describe the
evolution of black holes and the collapse of matter stars without
the central singularity, in this paper we continue to improve the
regular black holes proposed in \cite{Ling:2021olm} based on the
following consideration. Firstly, at Planck scale, the effects of quantum gravity are essentially strong to change the quantum mechanical behavior of matter, which  is usually reflected by generalized uncertainty principle(GUP). It is well known that GUP provides a lower bound for the size of any quantum object, whose size should be larger than the minimal length at Planck scale.  As  a result, when one puts any quantum object with finite size into a curved spacetime, it must be affected by the gravitational tidal force. In \cite{Ling:2021olm} , we presented an alternative point of view on this picture. One can assume that quantum object still obeys the usual Heisenberg's uncertainty relation, namely the quantum theory is retained, but introduce an effective gravitational field strength to count in the interacting effects of gravity and the quantum object. It turns out that in latter point of view, the black hole background will be modified to have a metric characterized by an effective Newton constant, leading to the regular black holes constructed in \cite{Ling:2021olm}.  Or in a word, the
regular black holes in \cite{Ling:2021olm}  are constructed based on the consideration of strong quantum gravity effects such as GUP. In this paper,
we expect that the effective
metric does not only contain the strong quantum gravity effects,
which may come from the GUP, but
also the 1-loop quantum corrections to the Newton potential via
effective field theory\cite{Bjerrum-Bohr:2002gqz,Donoghue:2012zc}.
Secondly, we expect the effective metric allows the time delay
between an observer at the central core and an observer at
infinity. This should be a more practical setup since any clock in
a gravitational potential well should be slowed down  in
comparison with the clock in an asymptotically flat region.  The
quite similar modification has been performed to the Hayward black
hole in \cite{DeLorenzo:2014pta}, here we will closely follow up
and improve their strategy to general regular black holes with
spherical symmetry. In particular, we will compare the distinct
behavior of Kretschmann curvature for these two sorts of black
holes.

This paper is organized as follows. In next section we will
present the general metric form for regular black holes with
spherical symmetry. Leave the basic features of these regular black
holes reviewed in Appendix,  we will focus on the mass dependent behavior of Kretschmann curvature when 1-loop quantum correction and time delay are taken into account. We will propose a scheme to figure
out the maximal time delay at the center by scanning the mass of
black holes such that the sub-Plackian feature of Kretschmann
curvature is preserved for black holes during the whole
evaporation process. Then from section three to
section five we will numerically demonstrate the dependent
behavior of Kretschmann curvature on various parameters in
various regular black holes. The distinct behavior are
compared for regular black holes with asymptotically Minkowski
core and those with asymptotically de-Sitter core. In particular,
we will find that Kretschmann curvature can always be
sub-Planckian, irrespective of the mass of black holes. Finally, we propose a scheme to fix the time delay parameter in the section of conclusion and discussion.

\section{The general setup for static spherically symmetric black hole }
We consider a static spherically symmetric black hole with a
general form of the metric
\begin{equation}\label{Eq.metric}
d s^{2}=-G(r) F(r) d t^{2}+\frac{1}{F(r)} d r^{2}+r^{2} d
\Omega^{2},
\end{equation}
with
\begin{equation}
 F(r)=1+2 \phi(r),
\end{equation}
where $\phi(r)$ is understood as the gravitational potential. In
general, we introduce $F(r)$ to include the strong quantum gravity
effects which would modify the singularity behavior at the Planck
scale, while $G(r)$ would be responsible for the weak quantum
gravity effects as well as the finite time delay between an
observer at the center and the one at infinity. Thus we assume
$G(r)$ is a regular function of radius, while the position of the
horizon $r_h$ is solely determined by $g^{rr}=F(r_h)=0$. For this
ansatz it is straightforward to derive the Kretschmann scalar
curvature which looks complicated and we present its expression in
Appendix.\ref{AppxA}.

In this paper we consider a specifical modification of the regular
black hole proposed by \cite{Ling:2021olm} closely following the
strategy presented in \cite{DeLorenzo:2014pta}. Then the
gravitational potential $\phi(r)$ and the modified function $G(r)$
are specified as
\begin{equation}\label{Eq.pG}
\phi=-\frac{M}{r} e^{-\alpha M^x / r^{n}}, \ \ \ \ \
G(r)=1-\frac{\beta M \gamma}{\gamma r^{3}+\beta M},
\end{equation}
where $\alpha$, $\beta$  and $\gamma$ are understood as
dimensionless constants. We have also set $l_p =M_p=1$ throughout
the paper. It implies that to recover the correction dimension of
any physical quantity, the unit $l_p$ or $m_p$ should be inserted
appropriately. The above form of $\phi(r)$ was originally proposed by us in \cite{Ling:2021olm}.  When $x$ and $n$ are specified with $n>x\geq 0$ and $n\geq 1$ , it produces various regular black holes with an asymptotically Minkowski core.  With $G(r)=1$, the sub-Planckian feature of  Kretschmann curvature as well as the thermodynamical behavior has been investigated in  \cite{Ling:2021olm}  as well, we refer to that paper for details.

The above form of $G(r)$ originally appears in
\cite{DeLorenzo:2014pta}, giving rise to a modified Hayward black
hole. Here we adopt the same form in order to introduce the 1-loop
quantum correction to the gravity potential and a time delay
between the center and infinity. Obviously, if $G(r)=1$, then it
goes back to the regular black hole proposed in
Ref.\cite{Ling:2021olm}. Now with non-trivial $G(r)$, it is also
straightforward to obtain the location of the outer horizon and
the thermodynamics of the above modified black holes, which are
presented in Appendix.\ref{AppxB} and Appendix.\ref{AppxC},
respectively. Here we intend to demonstrate how a one-loop quantum
correction and the time delay can be incorporated  by the function
$G(r)$. For this modified regular black holes, the metric
component $g_{00}$ at large scale behaves as
\begin{equation}
\begin{aligned}
g_{00}=1-\frac{2 M}{r}-\frac{\beta M }{r^3}+\frac{2\beta M^2 }{r^4}+\frac{2\alpha M^{x+1} }{r^{n+1}}+o\left(\frac{1}{r^{n+2}}\right).
\end{aligned}
\end{equation}

On the other hand, the leading quantum correction to the Newtonian
potential has been perturbatively computed and the large scale
behavior of gravitational potential takes the form as
\cite{Bjerrum-Bohr:2002gqz,Donoghue:2012zc}
\begin{equation}
\phi(r)=-\frac{M}{r}\left(1+
\frac{41}{10\pi}\frac{1}{r^2}+...\right),
\end{equation}
where the leading term is featured by a positive sign.  In this
paper, we require that $n\geq x \geq n/3$ and $n\geq 2$ such that
Kretschmann curvature can be sub-Planckian irrespective of the
mass of black holes, as analyzed in \cite{Ling:2021olm}. Therefore
the exponential form of the gravitational potential has no
contribution to the 1-loop quantum correction, we simply set
$\beta=41/(5 \pi)$ to reproduce such a 1-loop quantum
corrections\footnote{it is worthwhile to point our if  $x=0$ and
$n=2$, then the exponential form of the gravitational potential
would contribute a term with $\frac{2\alpha M }{r^3}$, then one
would set $\beta-2\alpha=41/(5 \pi)$.}.

Next we explain how the time delay is incorporated by $G(r)$ with
parameter $\gamma$. First of all, we remark that the metric
component $g_{00}$ at the center remains time-like which is in
contrast to the standard Schwarzschild black hole, thus we may
compare the time for two clocks which are placed at the center and
at infinity, respectively. Specifically, the time delay may be
defined as $(\delta t_{\infty}-\delta t_{0})/\delta
t_{\infty}=1-\sqrt{|g_{00}(0)}|$ as proposed
in\cite{DeLorenzo:2014pta}. Before introducing the function
$G(r)$, it is easy to see that $F(r)\rightarrow 1$ as
$r\rightarrow 0$, which means there is no time delay between these
two clocks, which looks peculiar since we know usually the
distribution of matter would lead to a time delay for any star or
matter collapse. Therefore we introduce a parameter $\gamma$ with
$1> \gamma \geq 0$ in $G(r)$ to produce a desired time delay at
the center. Obviously, now as $r\rightarrow 0$, $G(r)\rightarrow
1-\gamma$ and $F(r)\rightarrow 1$, then the request of time delay
between two clocks at the center and at infinity is measured by
$\gamma$.

\begin{figure} [t]
  \center{
  \includegraphics[scale=0.4]{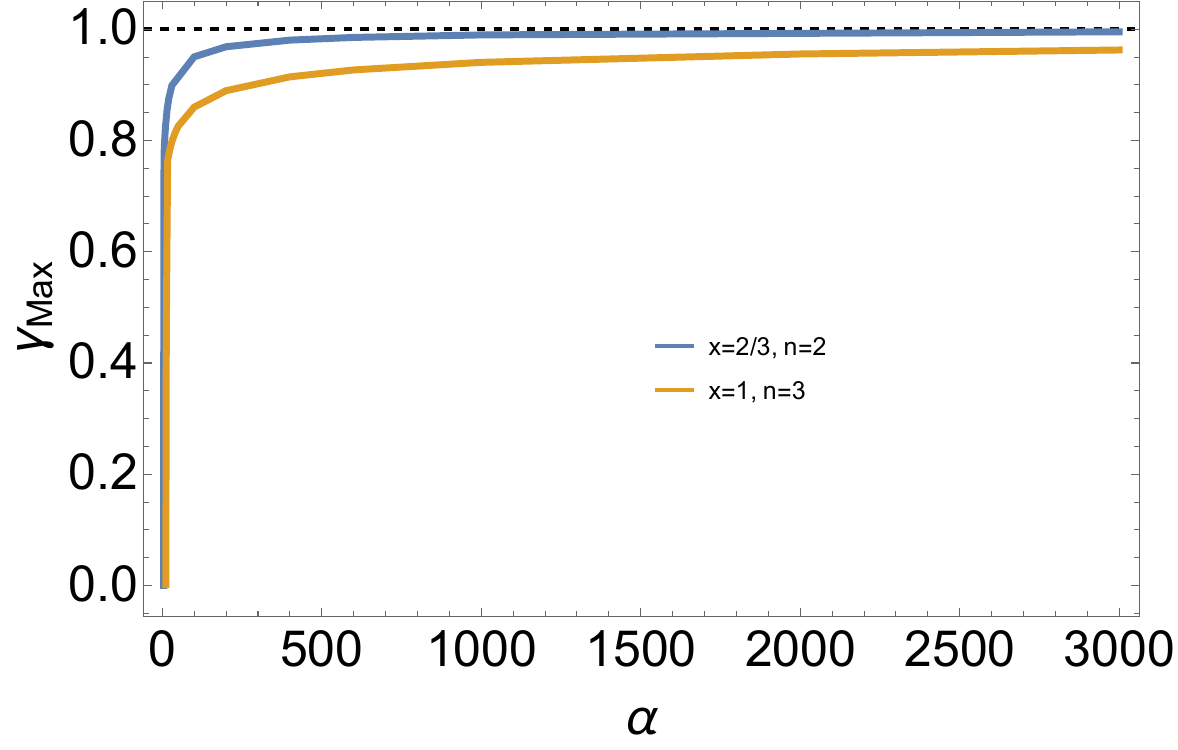}\ \hspace{0.05cm}
  \includegraphics[scale=0.4]{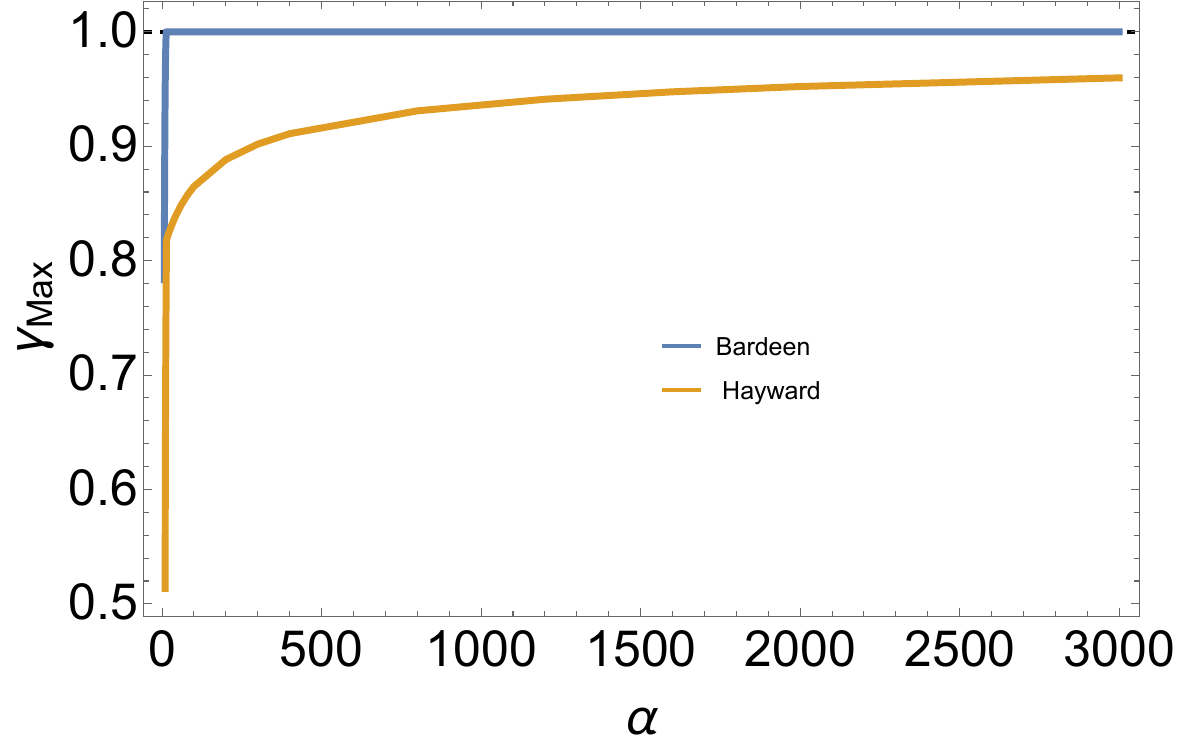}\ \hspace{0.05cm}
  \caption{\label{fig1v2} $\gamma_{max}$ as the function of $\alpha$, which is defined as $K^{max}(
\gamma_{max})=1$ and obtained by scanning the mass of black holes.
The left plot is for the regular black holes with asymptotically
Minkowski core while the right plot is for the regular black holes
with asymptotically de-Sitter core.}
  }
\end{figure}

Now we are concerned with the effects of $G(r)$ on Kretschmann
scalar curvature. As we have found in \cite{Ling:2021olm} where
$G(r)=1$, if $n\geq x \geq n/3$ and $n\geq 2$, then Kretschmann
curvature can be always sub-Planckian because its maximal value
$K_{max}$ is inversely proportional to the mass of black holes.
The saturated case is reached at $x=n/3$, where the maximal value
$K_{max}$ is independent of the mass of black holes. Specially,
when $n=2$ and $n=3$ the regular black holes with asymptotically
Minkowski core correspond to the Bardeen black hole and Hayward
black hole respectively, in the sense that they have the same
asymptotical behavior at large scales. Now, once the function
$G(r)$ is introduced, we point out that in general the maximal
value of Kretschmann curvature becomes a function of the mass $M$,
parameter $\alpha$ as well as the parameter $\gamma$, namely
$K^{max}(m,\alpha, \gamma)$. In particular, in this form as
$\gamma \rightarrow 1$, then the metric component $g_{tt}$ becomes
vanishing such that Kretschmann scalar curvature can easily exceed
the Planckian mass density, as we will explicitly demonstrate in
next sections. Of course this is not surprising, it just implies
that an arbitrarily large time delay is not available, which is
quite reasonable from the physical side because any time delay
induced by the distribution of matter should be finitely large.
Therefore, once $\alpha$ is given, we intend to define a maximal
value of $\gamma$ that saturates the bound of Kretschmann scalar
curvature by scanning the mass of black holes, namely, $K^{max}(
\gamma_{max})=1$ for some certain mass $m$. More importantly,  we
remark that in the region of $n\geq x \geq n/3$ and $n\geq 2$,
this can always be done because in this region $K^{max}$ will not
increase with the mass $M$ forever, but becoming saturated at
large $M$ of black holes. This will be justified by the numerical
analysis as well in next sections. Therefore, for a given
$\alpha$, we can obtain the maximal time delay $\gamma_{max}$ such
that given a $\gamma$ under the condition $\gamma_{max}>\gamma\geq
0$, Kretschmann curvature will maintain at the sub-Planckian scale
for all the masses of a black hole. This of course is what we
expect because once all the parameters are specified in
Eq.(\ref{Eq.pG}), we hope the sub-Planckian feature of $K$ is
preserved during the whole
 evaporation process, in which the mass of black hole changes. Therefore, we numerically plot $\gamma_{max}$ as the
function of $\alpha$ for some typical regular black holes in
Fig.\ref{fig1v2}. It is obvious to see that in general
$\gamma_{max}$ grows up rapidly with the increase of parameter
$\alpha$ and then becomes saturated as $\gamma_{max}\rightarrow
1$. In particular, for Bardeen black hole it approaches one
quickly in the region with smaller $\alpha$. Anyway, for all the
black holes below we will consider the time delay with
$\gamma_{max}>\gamma\geq 0$ such that the sub-Planckian feature is
always guaranteed for Kretschmann scalar curvature.

We argue that our above treatment is a dramatic improvement in
comparison with  the scheme  applied in \cite{DeLorenzo:2014pta},
where the maximal value of $\gamma$ is defined for a given mass.
Because in next sections will find that for a given $\alpha$ and
$\gamma$, Kretschmann curvature is sub-Planckian with some large
mass does {\it not}  guarantee that it must be sub-Planckian with
any mass, even if it becomes saturated in large mass limit (this
is always true indeed). On the contrary, Kretschmann curvature may
exceed the Planck scale easily when the mass drops down. For
instance, in Figure 6 of \cite{DeLorenzo:2014pta} which is on the
modified Hayward black hole,  $K$ is sub-Planckian for $M=10^5$,
but if one drops down the mass with other parameters fixed,
$K_{max}$ can easily exceed one. For instance, if $M=10^2$, then
$K_{max}\simeq 6000$. In a word, the introduction  of
$\gamma_{max}$ {\it independent of mass} in our paper gives us a
way to specify the parameter values such that the sub-Planckian
feature of curvature can be preserved during the whole process of
evaporation.

\section{Modified regular black hole with $x=1$ and $n=2$}

\begin{figure} [t]
  \center{
  \includegraphics[scale=0.33]{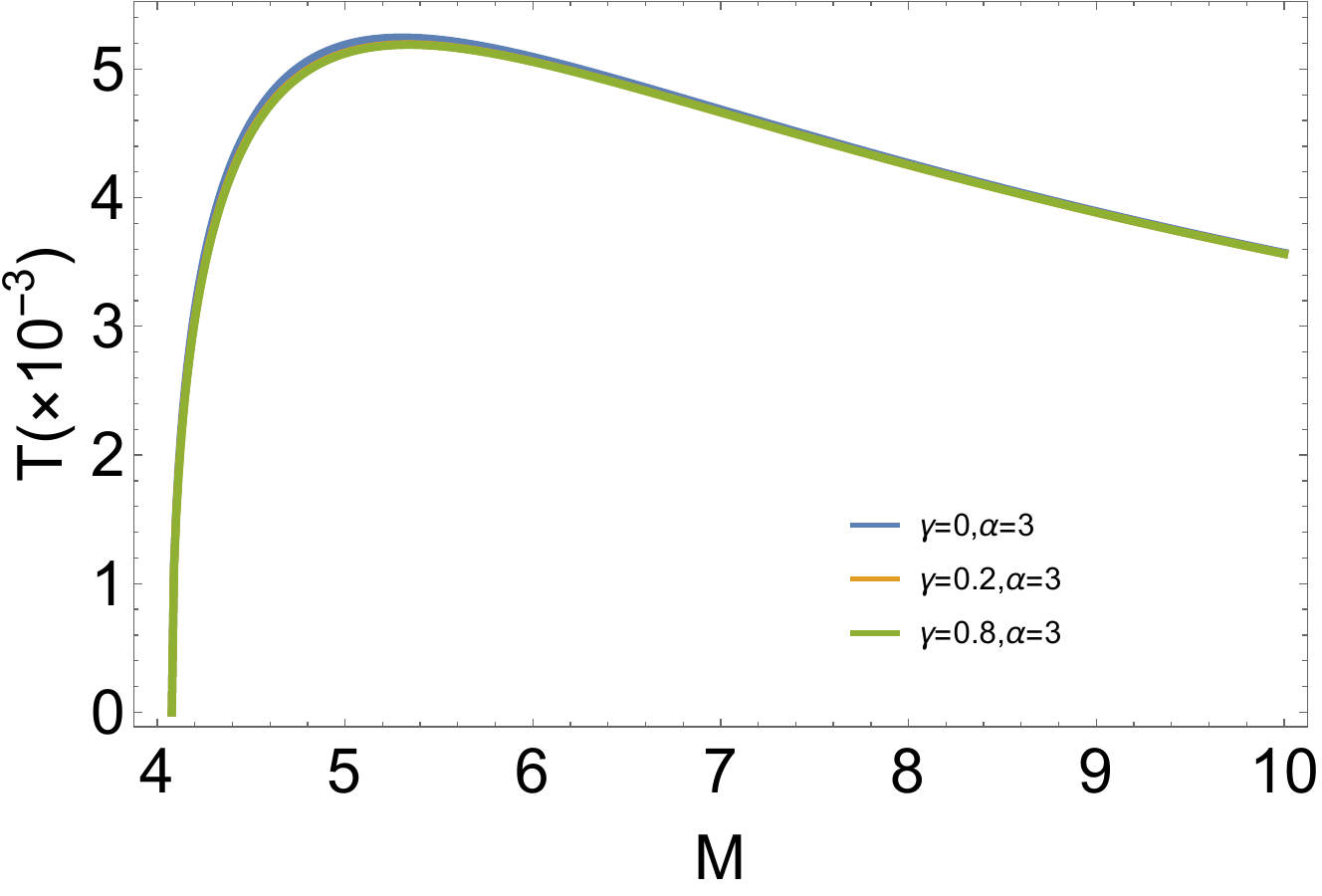}\ \hspace{0.05cm}
  \includegraphics[scale=0.37]{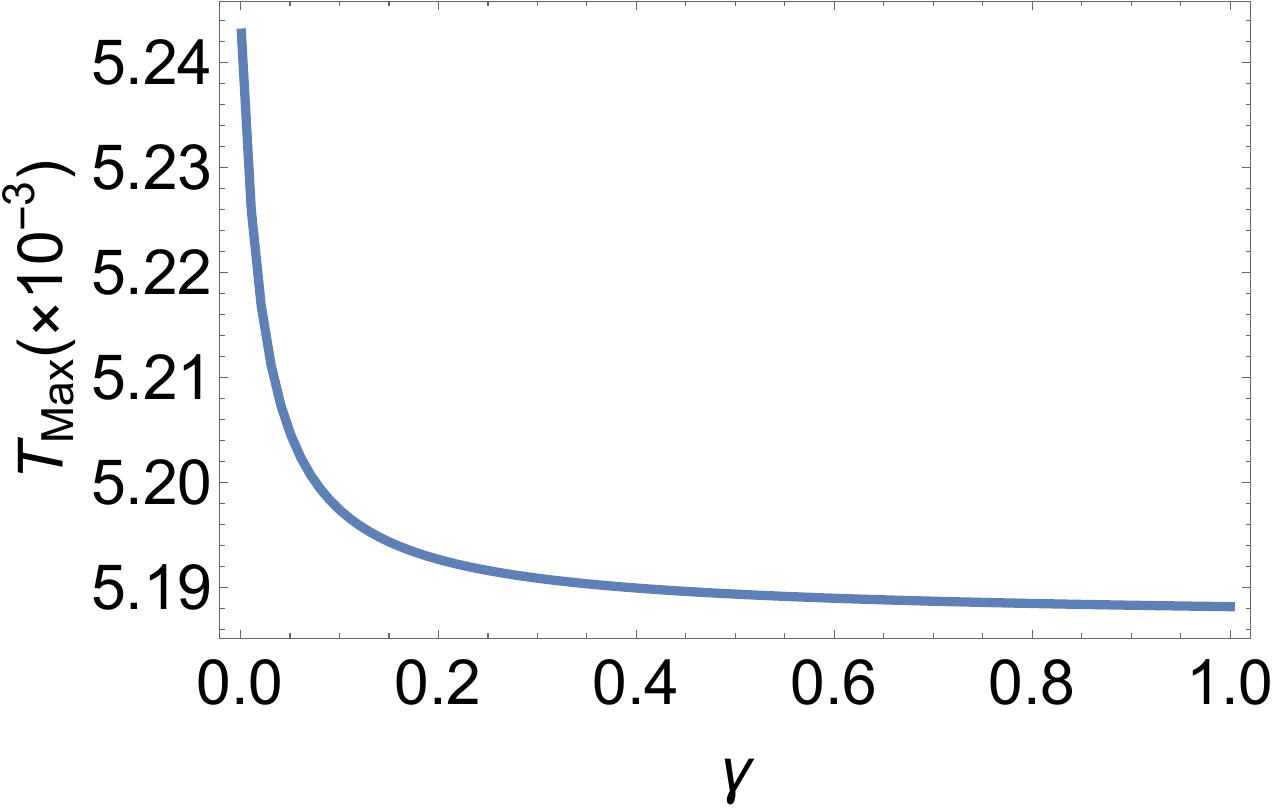}\ \hspace{0.05cm}
  \caption{\label{fig3} Left: The temperature  as the function  of $M$. \ \ Right: The maximum value of temperature  as the function  of $\gamma$. }}
\end{figure}

\begin{figure} [t]
  \center{
  \includegraphics[scale=0.39]{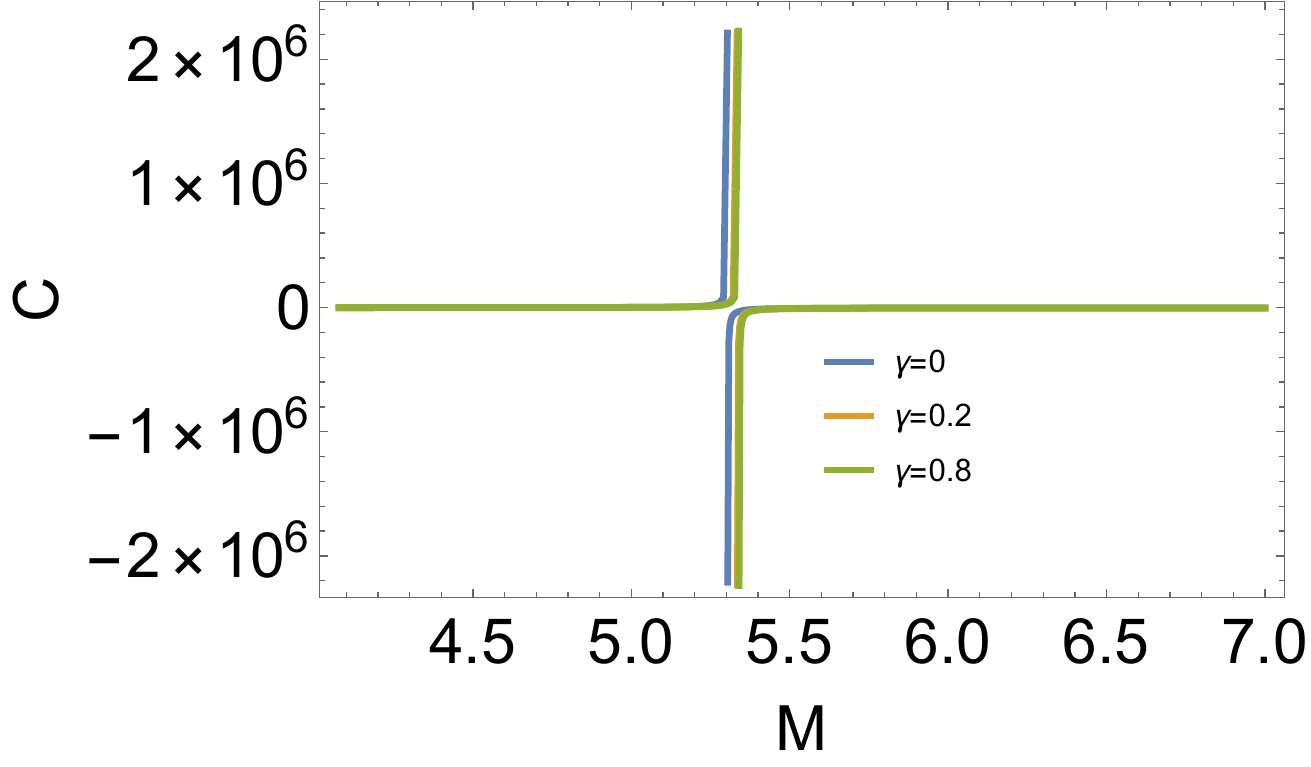}\ \hspace{0.05cm}
  \includegraphics[scale=0.34]{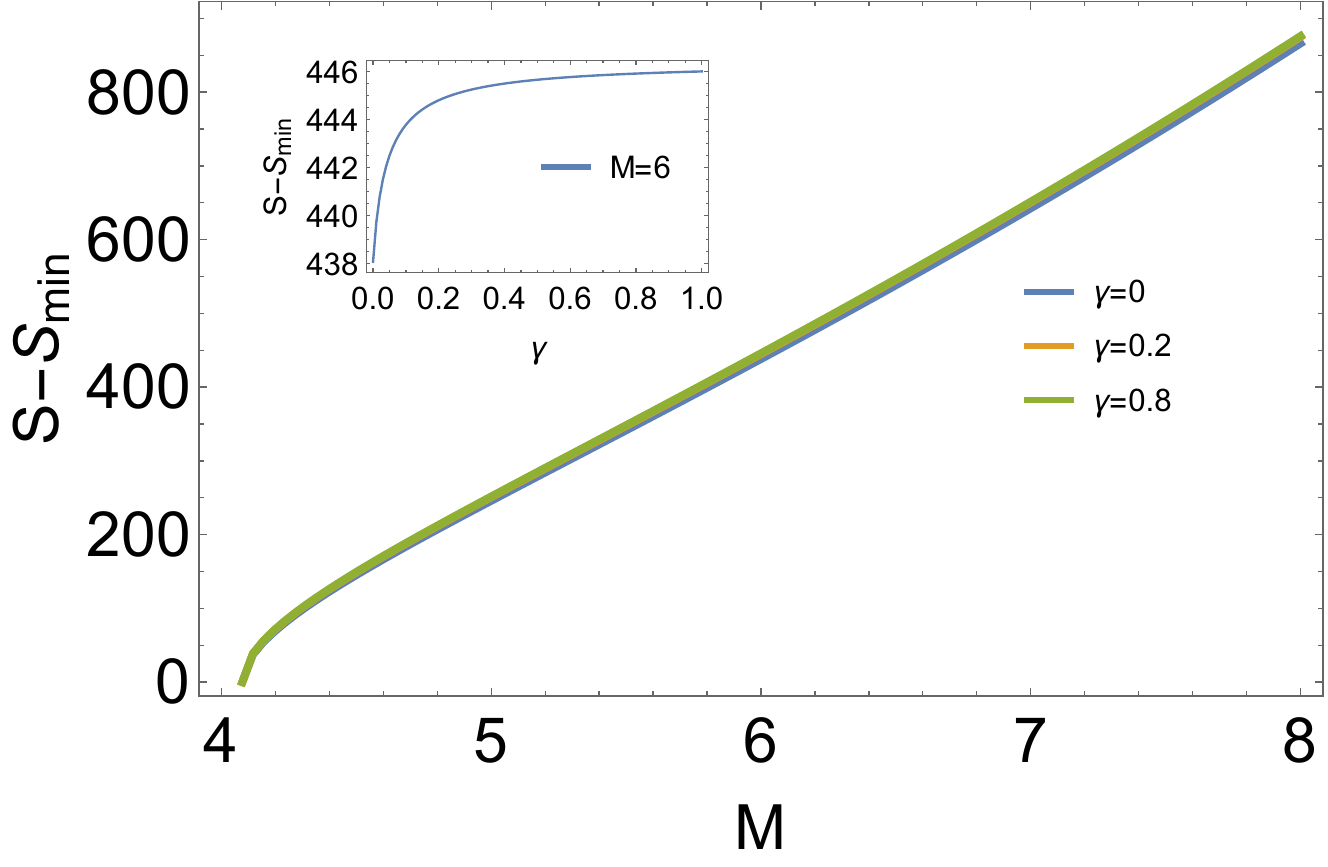}\ \hspace{0.05cm}
  \caption{\label{fig3v2} Left: The heat capacity  as the function  of $M$. \ \ Right:  The entropy  as the function  of $M$, while the inset shows the entropy as the function of $\gamma$ with a given mass. }}
\end{figure}

In this section  we study the modified regular black hole with
$n=2$ and $x=1$. When $G(r)=1$, its main features have been
investigated in \cite{Ling:2021olm}.  We remark that the following
features are maintained  when the metric is modified by $G(r)$.
Firstly, the mass of the black hole is bounded by $M \geq \frac{e
\alpha}{2}$. In particular, when this bound is saturated, the
black hole is characterized by the minimal radius
$r_h=\sqrt{e}\alpha_0$, which may be treated as the remnant of the
black hole evaporation, since the final stage of the black hole is
characterized by $M\rightarrow \frac{e \alpha_{0}}{2}$ (or
    $r_h\rightarrow\sqrt{e}\alpha_0$). This effect completely results
    from the exponential suppressing potential and is controlled by
    the parameter $\alpha$. In this limit, it is obvious to see
that the Hawking  temperature is always vanishing even when 1-loop
quantum correction and time delay are incorporated, as illustrated
in Fig.\ref{fig3}.  It is noticed that the maximal value of the
temperature goes down with the increase of $\gamma$.
In addition, based on the equations in Appendix.\ref{AppxC}  we may plot the heat capacity as well as the entropy as the function of black hole mass. Whenever the temperature reaches the maximal value, the heat capacity becomes divergent, as  illustrated in Fig.\ref{fig3v2}, which means the black hole undergoes a transition from a system with negative heat capacity to a system with positive heat capacity during the process of evaporation.  Since the black hole has a remnant with the minimal mass, whose entropy may be denoted as $S_{min}$,  we may also plot the entropy difference between the black hole with any mass and the one with the minimal mass, as illustrated in Fig.\ref{fig3v2}. It is found that the entropy difference monotonically decreases with the decrease of the mass $M$, but does not change much with the change of parameter $\gamma$. The inset of Fig.\ref{fig3v2} tells us that the entropy increases a little bit when $\gamma$ goes up with a given mass.

\begin{figure} [t]
  \center{
  \includegraphics[scale=0.33]{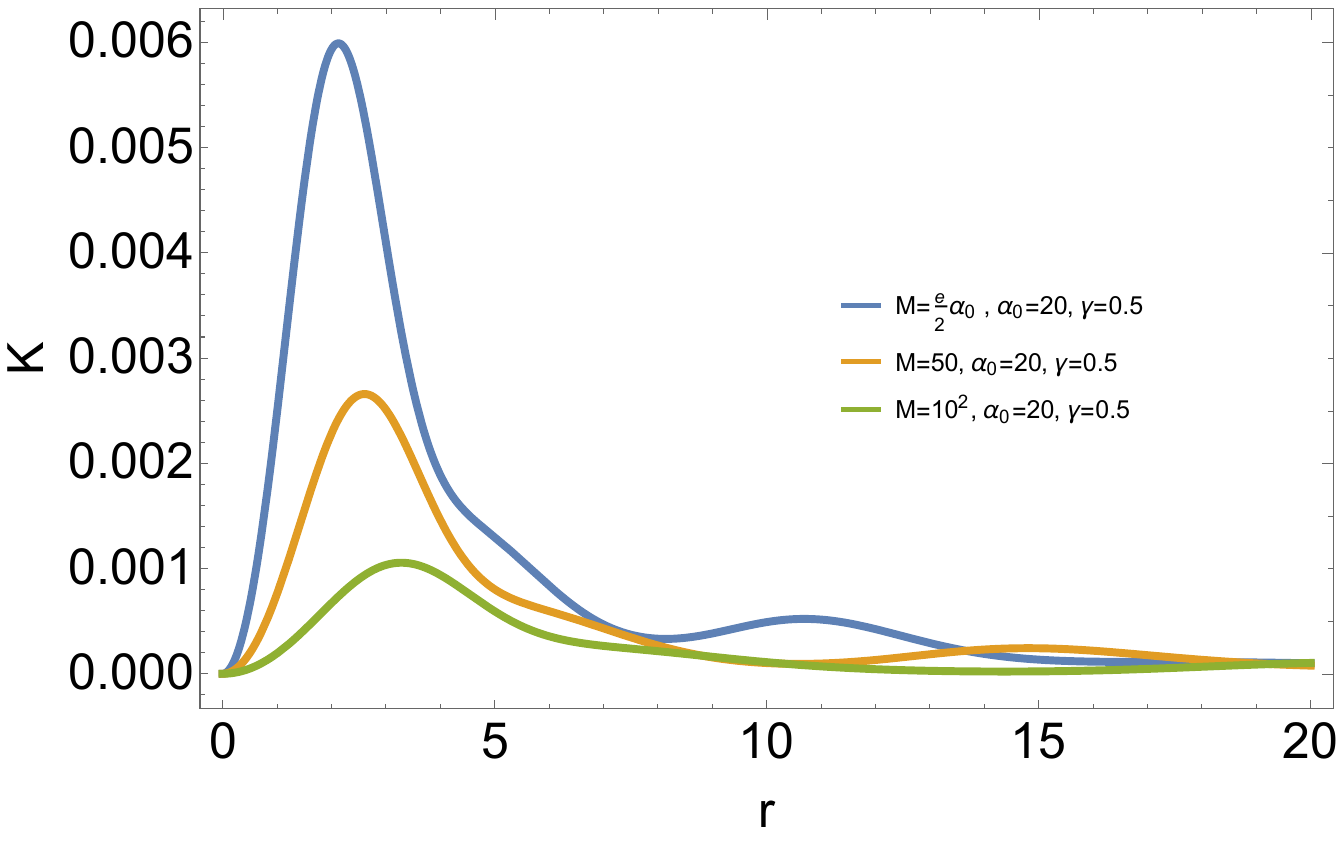}\ \hspace{0.05cm}
  \includegraphics[scale=0.31]{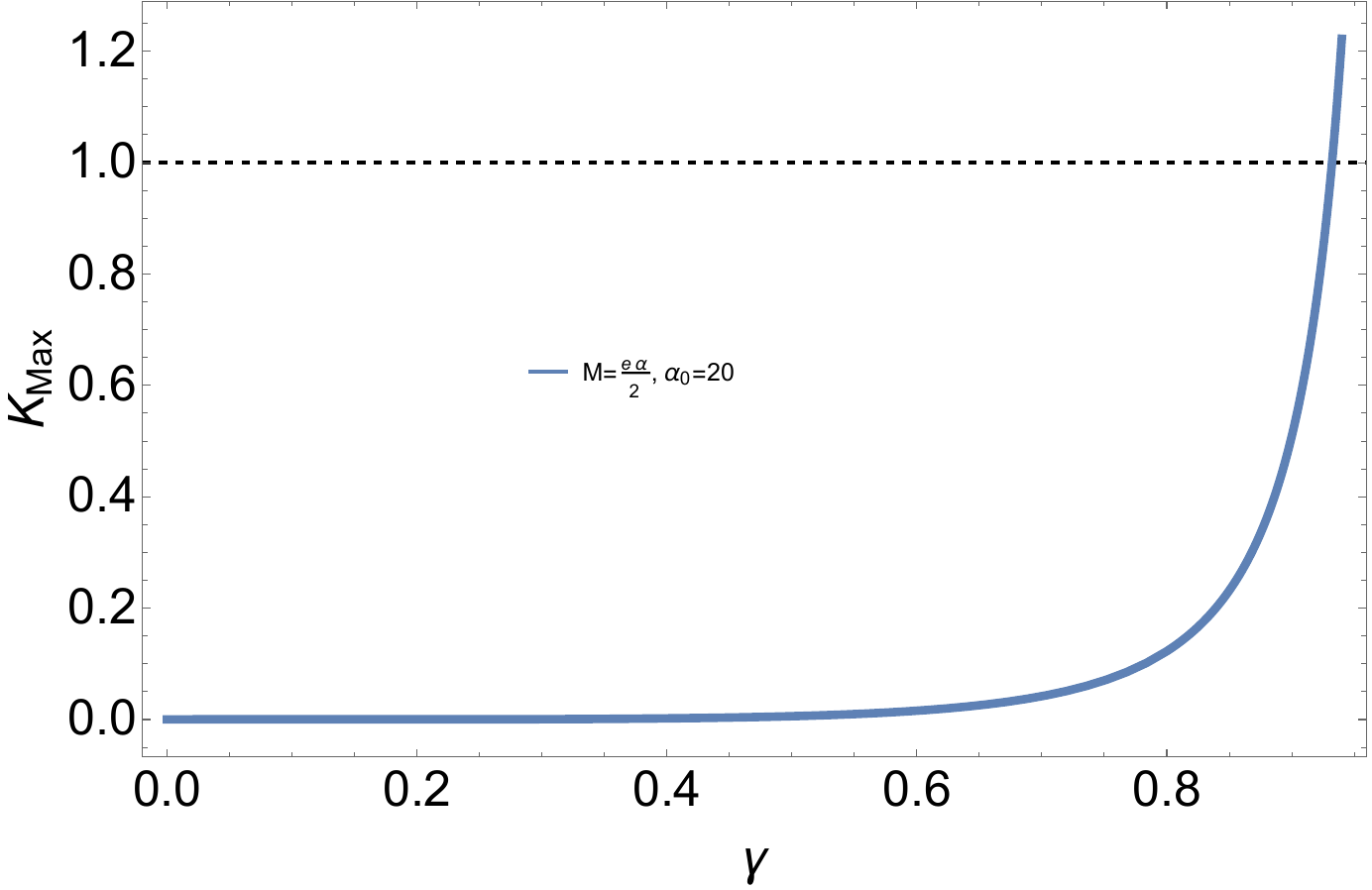}\ \hspace{0.05cm}
  \caption{\label{fig1}  Left: The Kretschmann scalar curvature $K$  as the function of the radial coordinate $r$ with $\gamma=0.5$.  Right:The maximum value of the Kretschmann scalar curvature $K$ as the function of  $\gamma$. }}
\end{figure}

Secondly, these exists a minimal value for $\alpha$ at the Planck
scale $\alpha_{min}\simeq 2.747$ such that
$K^{max}(\alpha_{min})=1$. When 1-loop quantum correction and time
delay are incorporated, we find $\alpha_{min}$ does not change.
For $\alpha\geq \alpha_{min}$, we may plot Kretschmann curvature
$K$  as the function of the radial coordinate $r$, as shown in
Fig.\ref{fig1}. In general, the maximal value of $K$ appears at
some position in space, and we denote it as $K_{max}$. Similar to
the case with $G(r)=1$,  we find $K(0)=0$ always, but the location
of $K_{max}$ moves to the right side with the increase of mass
$M$. We demonstrate $K_{max}$ as the function of $\gamma$ in the
right plot of Fig.\ref{fig1}, indicating that it monotonously
grows up with the parameter $\gamma$ once $\alpha$ is fixed.
Therefore, we need define a $\gamma_{max}$ to preserve Kretschmann
curvature to be sub-Planckian.

\begin{figure} [t]
  \center{
  \includegraphics[scale=0.4]{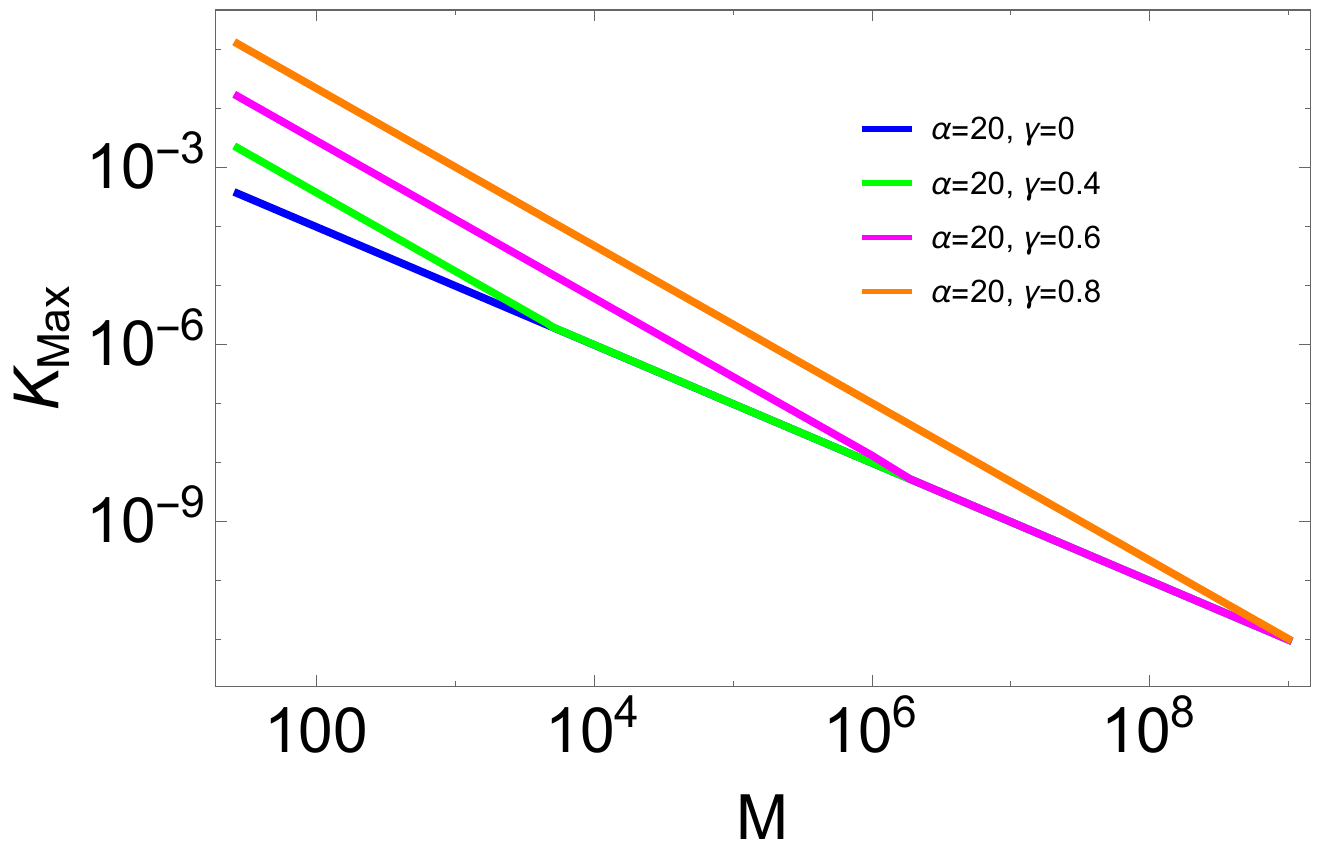}\ \hspace{0.05cm}
  \caption{\label{fig1vv2} The maximum value of the Kretschmann scalar curvature $K$ as the function of $M$. }}
\end{figure}

Finally, we illustrate the mass dependent behavior of $K_{max}$ in
Fig.\ref{fig1vv2}. For $G(r)=1$, we have figured out in
\cite{Ling:2021olm} that $K_{max}\propto 1/M$ such that the black
hole with the minimal mass is featured by the maximal $K_{max}$.
When $G(r)$ is turned on, we find in this figure that $K_{max}$
remains a linear relation with the logarithm of mass, thus
decaying with the increase of mass. The main difference is that
the value of $K_{max}$ is promoted due to the presence of
$\gamma$.

\section{Modified regular black hole with $x=2/3$ and $n=2$}

In this section, we will study the modified regular black hole
with $n=2$ and $x=2/3$,  which corresponds to Bardeen black hole
at large scales, since the gravitational potential of Bardeen
black hole $\phi(r)$ is given as
\begin{equation}
\phi(r)=-\frac{M r^{2}}{\left(\frac{2}{3} \alpha M^{2/3}+r^{2}\right)^{3 / 2}}.
\end{equation}
One can easily check these two black holes have the same expansion
behavior for large radius. Nevertheless, they have distinct
behavior near the central core. The former has an asymptotically
Minkowski core, while Bardeen black hole has a de-Sitter core.

Similarly, we point out that for the regular black hole with
$x=2/3, \ n=2$, the mass is bounded by $M
\geq\left(\frac{e}{2}\right)^{3 / 4} \alpha^{3 /4}$, which is
solely determined by the exponentially suppressing potential
controlled by the parameter $\alpha$. Thus the presence of $G(r)$
does not change the whole picture of black hole evaporation, and
the remnant of black holes maintains at the final stage. Again,
$\alpha$ has the minimal value $\alpha_{min} \simeq 0.875$, that
is defined by $K_{max}(\alpha_{min})=1$ with $\gamma=0$. While for
Bardeen black hole, it is worthwhile to point out that $M
\geq(\frac{9 \alpha }{4})^3$, indicating that $M_{min}$ increases
with $\alpha^3$. That is to say, $M_{min}$ grows rapidly with
$\alpha>1$. This behavior is in contrast to all the other regular
black holes in this paper.
\begin{figure} [t]
  \center{
  \includegraphics[scale=0.35]{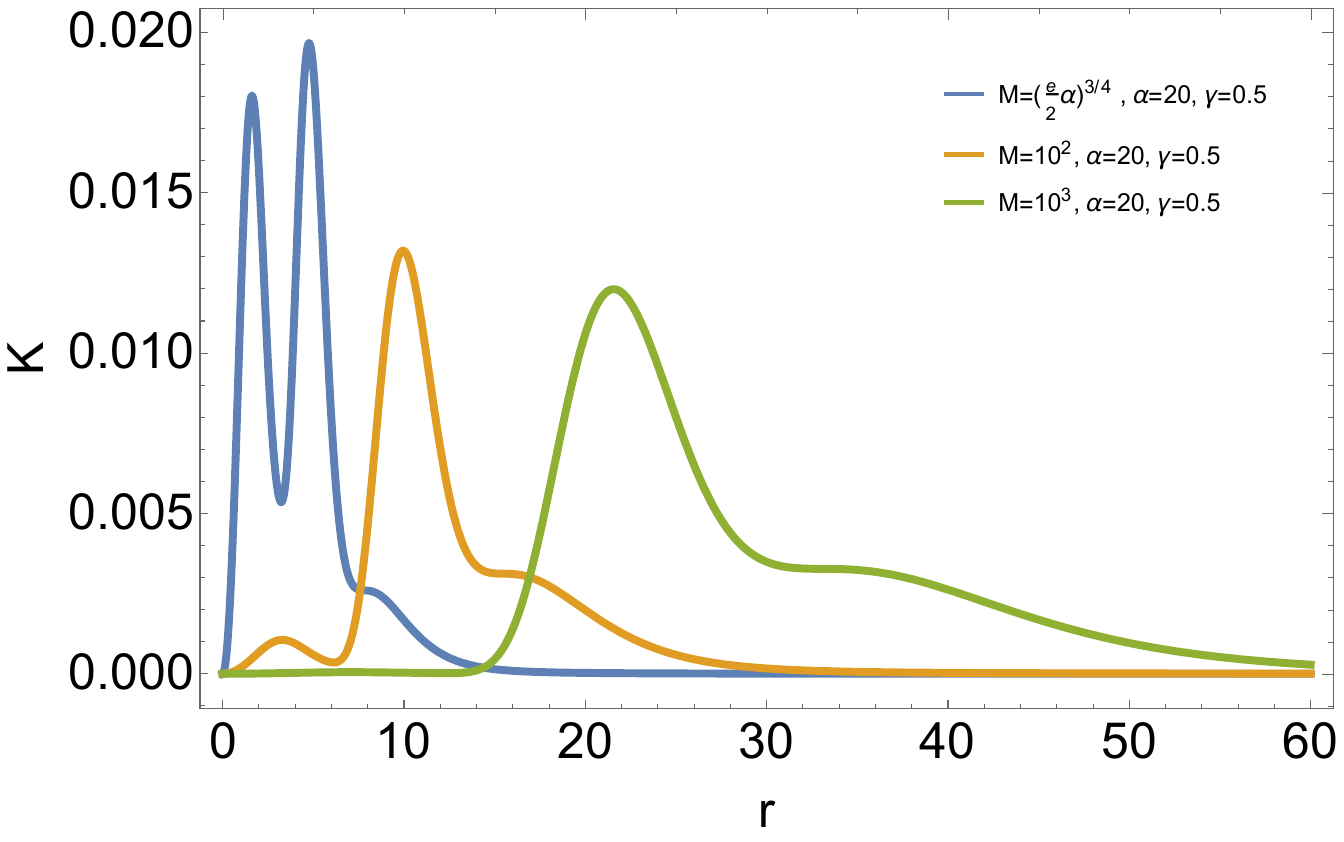}\ \hspace{0.05cm}
  \includegraphics[scale=0.35]{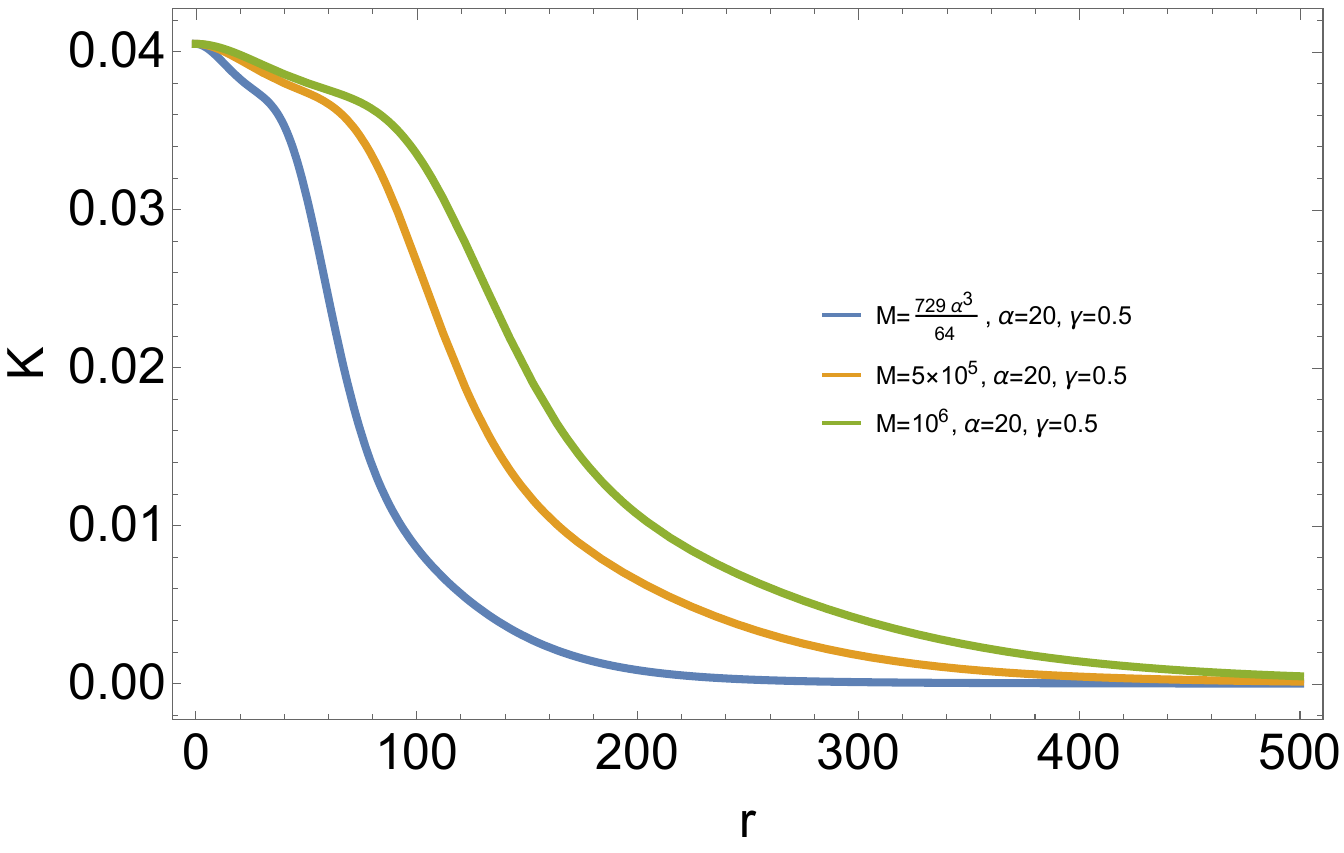}\ \hspace{0.05cm}
\caption{\label{fig4}  Kretschmann curvature $K$  as the function
of the radial coordinate $r$ for $\gamma=0.5$. The left plot is
for the regular black hole with $x=2/3, \ n=2$, while the right
plot is for Bardeen black hole.}}
\end{figure}

\begin{figure} [t]
  \center{
  \includegraphics[scale=0.28]{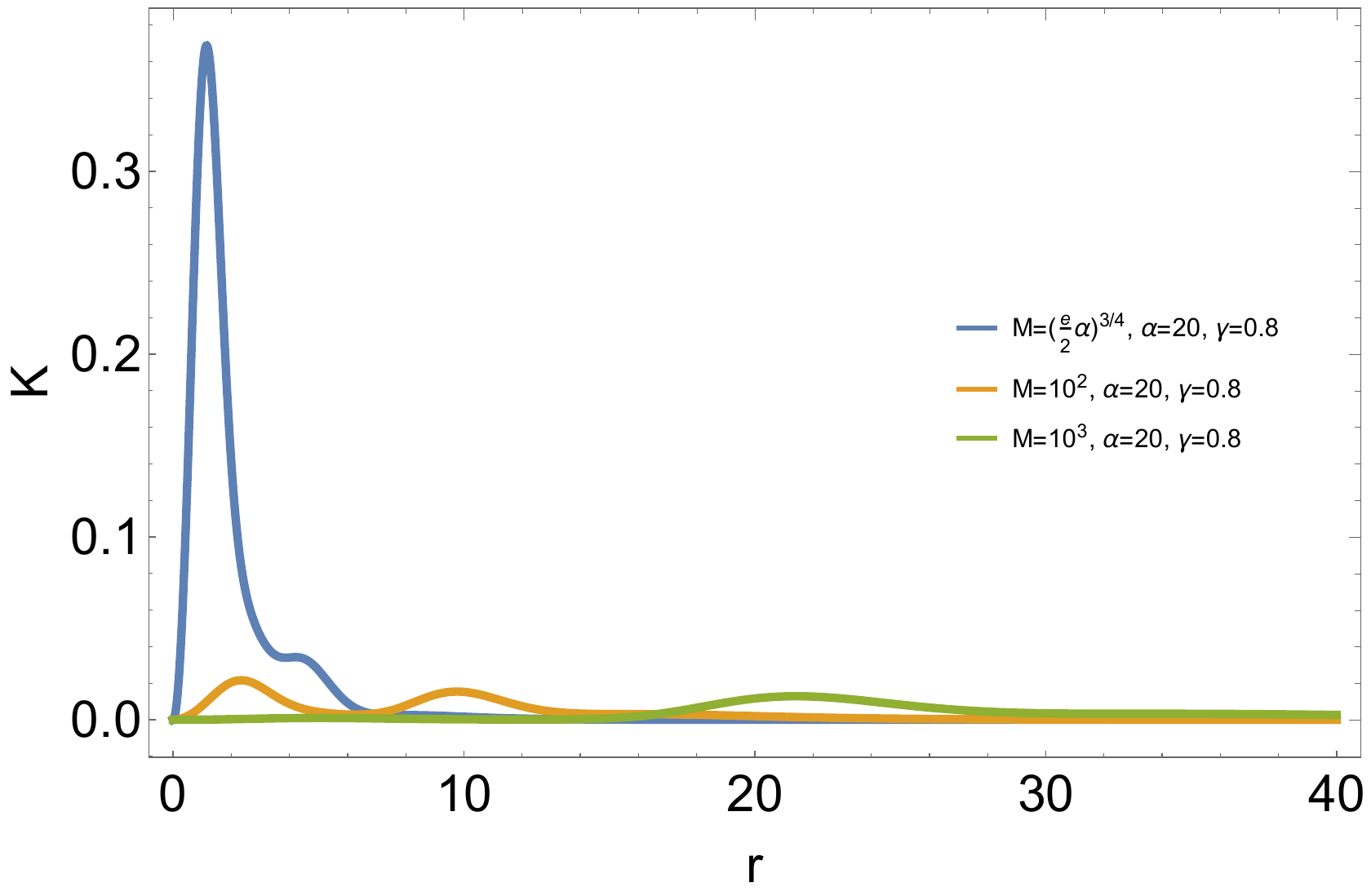}\ \hspace{0.05cm}
  \includegraphics[scale=0.28]{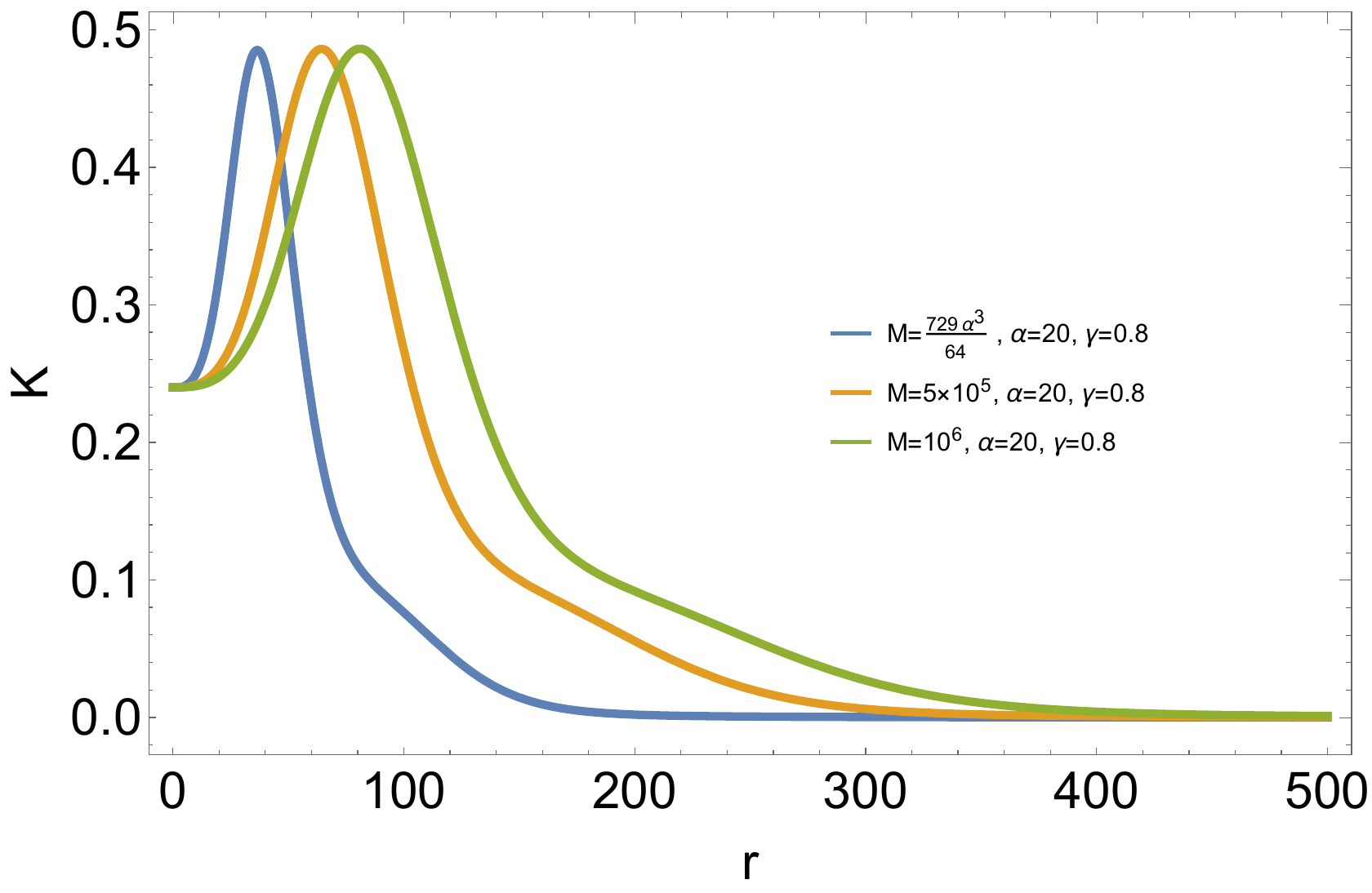}\ \hspace{0.05cm}
\caption{\label{fig4v3}  Kretschmann curvature $K$  as the
function of the radial coordinate $r$ for $\gamma=0.8$. The left
plot is for the regular black hole with $n=2, \ x=2/3$, while the
right plot is for Bardeen black hole.}}
\end{figure}

Now let us compare Kretschmann scalar curvature $K$  as the
function of the radius $r$ for these two black holes. In
Fig.\ref{fig4} we plot  Kretschmann  curvature $K(r)$ for
$\gamma=0.5$,   which is much smaller than $\gamma_{max}$. One
finds $K(0)=0$ always  for black hole with $x=2/3$ and $n=2$,
while $K(0)=K_{max}$ for Bardeen black hole, irrespective of the
mass of the black hole. While for $\gamma$ close to
$\gamma_{max}$, for instance $\gamma=0.8$ as illustrated in
Fig.\ref{fig4v3}, the maximal value of $K$ runs away from the
center for Bardeen black hole as well.

\begin{figure} [t]
  \center{
  \includegraphics[scale=0.351]{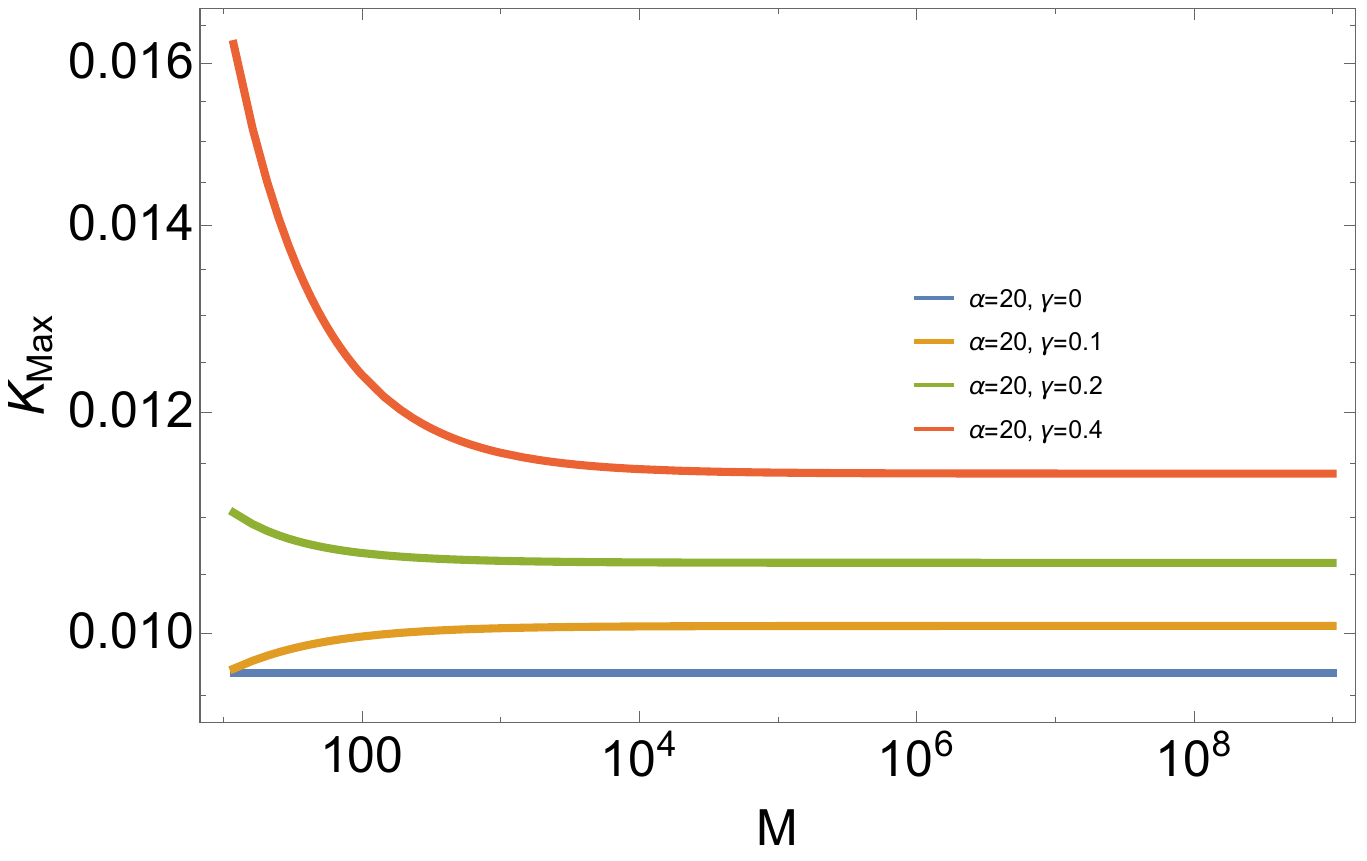}\ \hspace{0.05cm}
  \includegraphics[scale=0.345]{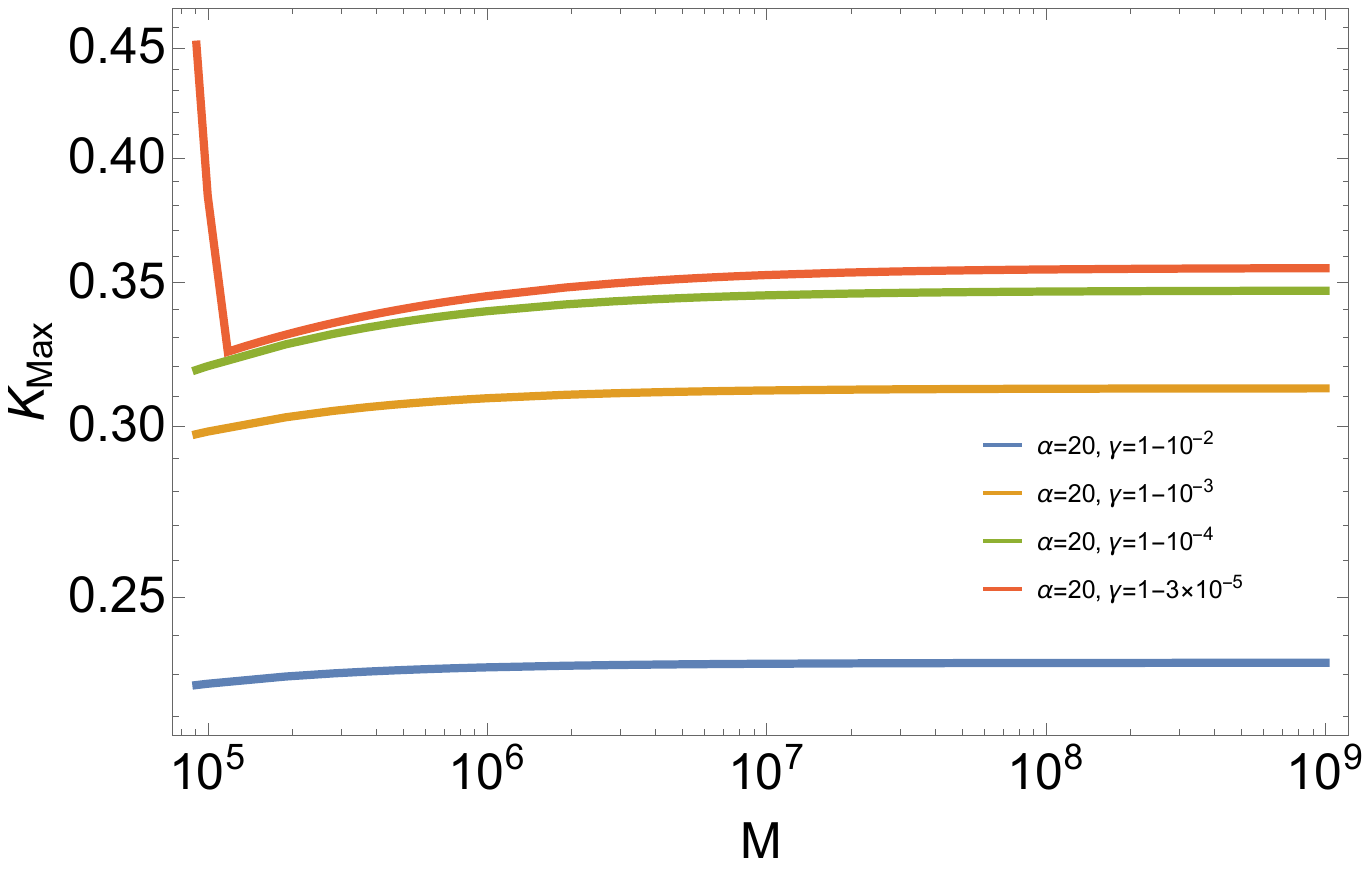}\ \hspace{0.05cm}
\caption{\label{fig4v2} The maximum value of Kretschmann curvature
$K_{max}$ as the function of $M$ for the black hole with $n=2, \
x=2/3$ (left) and Bardeen black hole (right).}}
\end{figure}

\begin{figure} [t]
  \center{
  \includegraphics[scale=0.345]{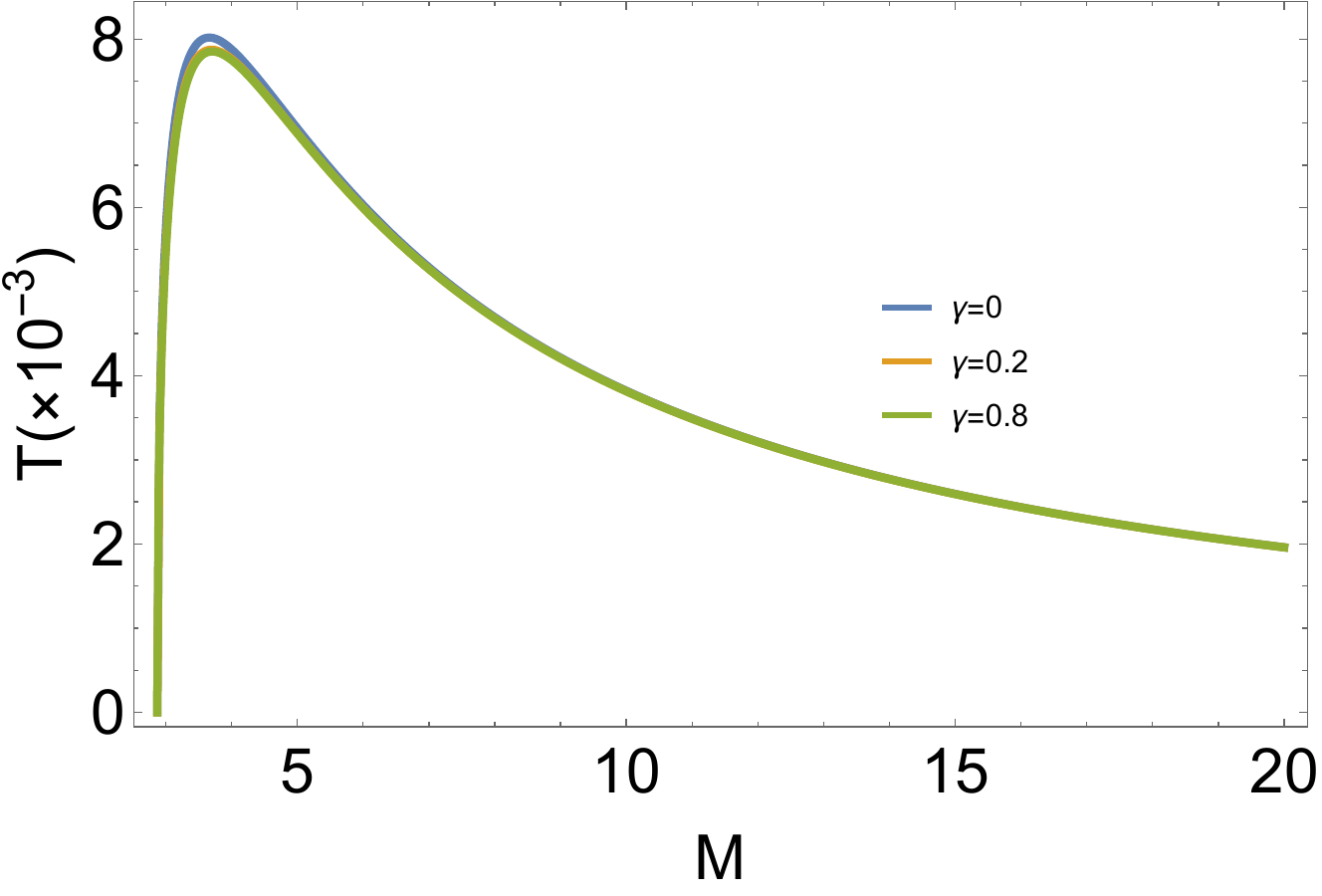}\ \hspace{0.05cm}
  \includegraphics[scale=0.38]{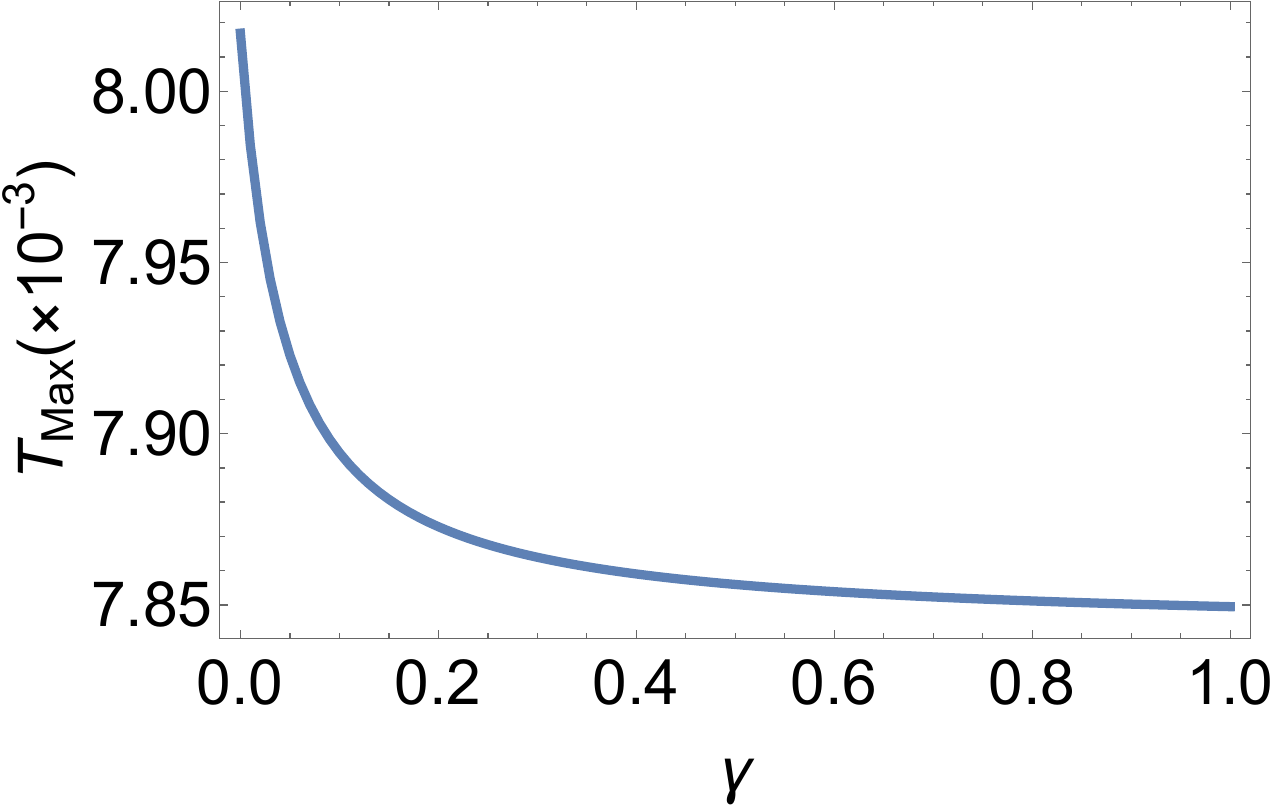}\ \hspace{0.05cm}
\caption{\label{fig6} Left: The temperature  as the functions  of
$M$. \ \ Right: The maximum value of temperature  as the functions
of $\gamma$. }}
\end{figure}

\begin{figure} [t]
  \center{
  \includegraphics[scale=0.39]{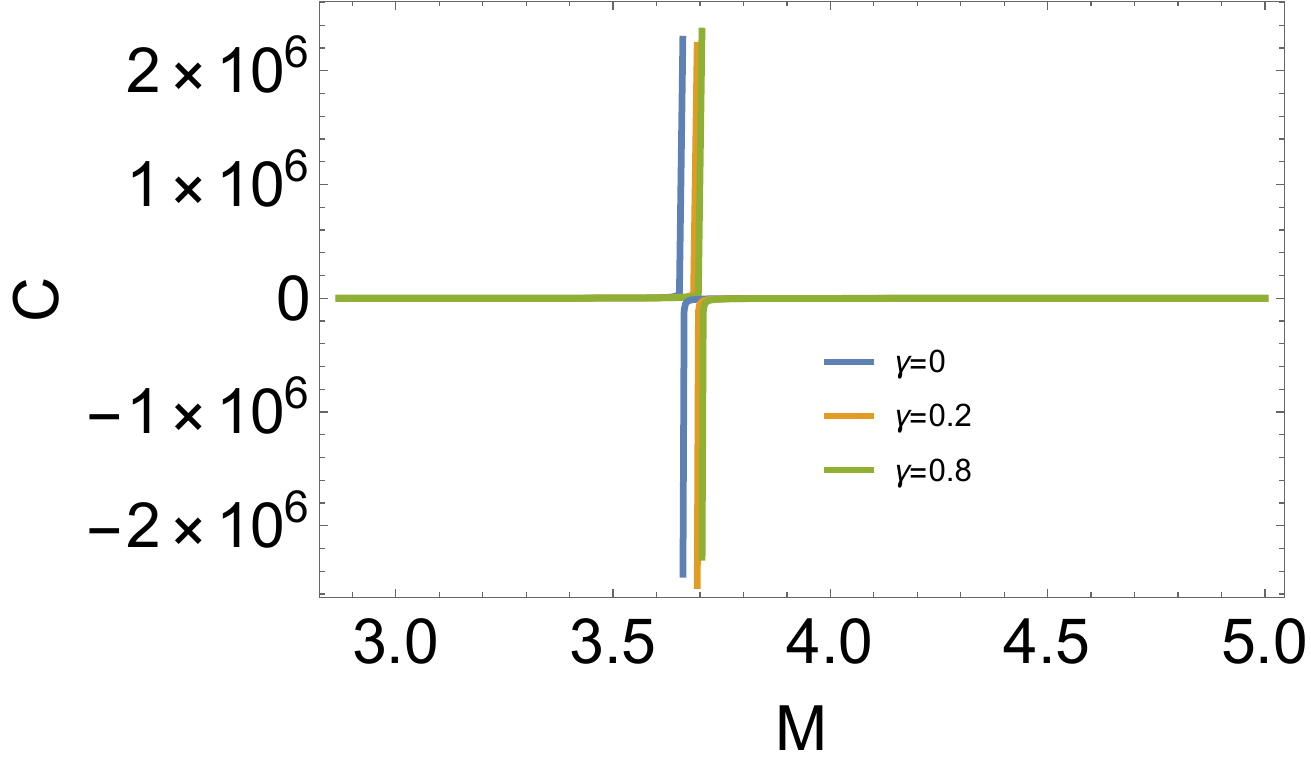}\ \hspace{0.05cm}
  \includegraphics[scale=0.35]{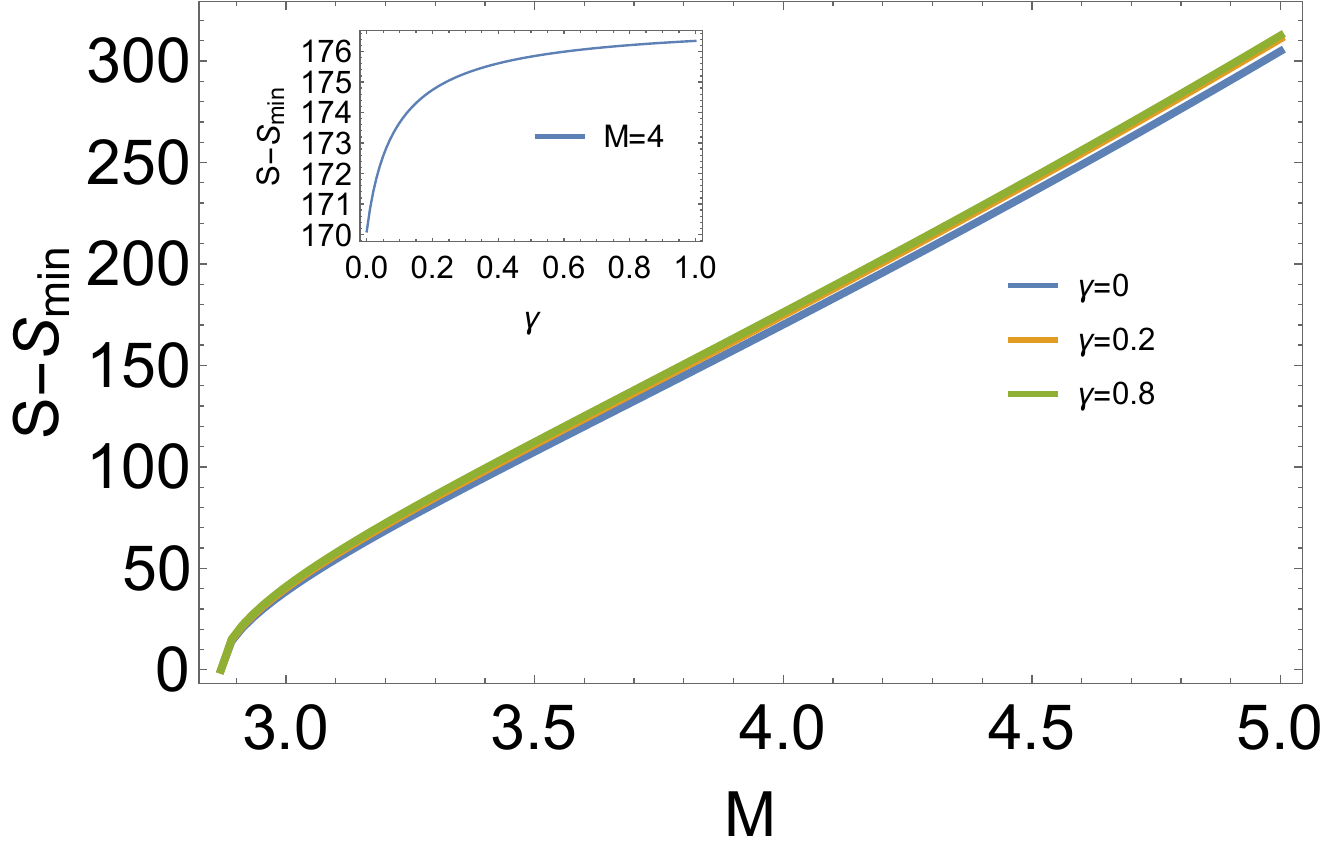}\ \hspace{0.05cm}
  \caption{\label{fig6v2} Left: The heat capacity  as the function  of $M$. \ \ Right:  The entropy  as the function  of $M$, while the inset shows the entropy as the function of $\gamma$ with a given mass. }}
\end{figure}

Next we focus on the mass dependent behavior of $K_{max}$ for
these two black holes. It is known in \cite{Ling:2021olm} that
when $G(r)=1$, both black holes are characterized by the fact that
$K_{max}$ is independent of the mass of black holes. However, when
$G(r)$ is turned on we find this feature does not hold anymore.
This is not surprising if one recalls that an arbitrarily large
time delay would make the metric singular again. Therefore, in
this case we demonstrate the mass dependent behavior of $K_{max}$
for different values of $\gamma$ in Fig.\ref{fig4v2}. First of
all, when $G(r)=1$, namely $\gamma=0$, $K_{max}$ is a constant
indeed (see the left plot of Fig.\ref{fig4v2}), while with the
increase of $\gamma$, we notice that $K_{max}$ becomes larger and
depends on the mass. In particular, it is surprising to notice
that as $\gamma$ approaches $\gamma_{max}$, $K_{max}$ with small
masses becomes larger, and this phenomenon is observed for both
black hole with $x=2/3, \ n=2$ and Bardeen black hole, as
illustrated in Fig.\ref{fig4v2} ($\gamma=0.4$ in the left plot
while $\gamma=1-3\times 10^{-5}$ in the right plot). It is this
fact that $K_{max}$ with small masses always grows up to one at
first and determines the value of $\gamma_{max}$ when we scan the
mass of black holes. This tendency also gives us more difficulty
to figure out $\gamma_{max}$ by scanning the mass of black holes
in Fig.\ref{fig1v2}. On the other hand, $K_{max}$ always becomes
saturated in large mass limit. This nice feature is important for
us to pick out a $\gamma_{max}$ to preserve $K_{max}$ to be
sub-Planckian always. In addition, we remark that for Bardeen
black hole, the mass dependent behavior of $K_{max}$  does not
change much until $\gamma\rightarrow \gamma_{max}$, and the mass
scale is much larger than that for the black hole with $x=2/3, \
n=2$, which is the reflection of the fact $M_{min}\propto
\alpha^3$ since we have set $\alpha=20$.

Finally,  the
thermodynamical behavior of the black hole is
illustrated in Fig.\ref{fig6} and Fig.\ref{fig6v2}.  We find that the
maximal value of the Hawking temperature becomes smaller slightly
with the increase of $\gamma$, while the entropy becomes larger slightly with the increase of $\gamma$.

\section{Modified regular black hole with $x=1$ and $n=3$}

In this section we will study the modified regular black hole with
$n=3$ and $x=1$ in a quite parallel manner, which corresponds to
Hayward black hole at large scales. The gravitational potential
for Hayward black hole is given by
\begin{equation}
\phi(r)=-\frac{M r^{2}}{r^{3}+M \alpha}.
\end{equation}

\begin{figure} [t]
  \center{
  \includegraphics[scale=0.35]{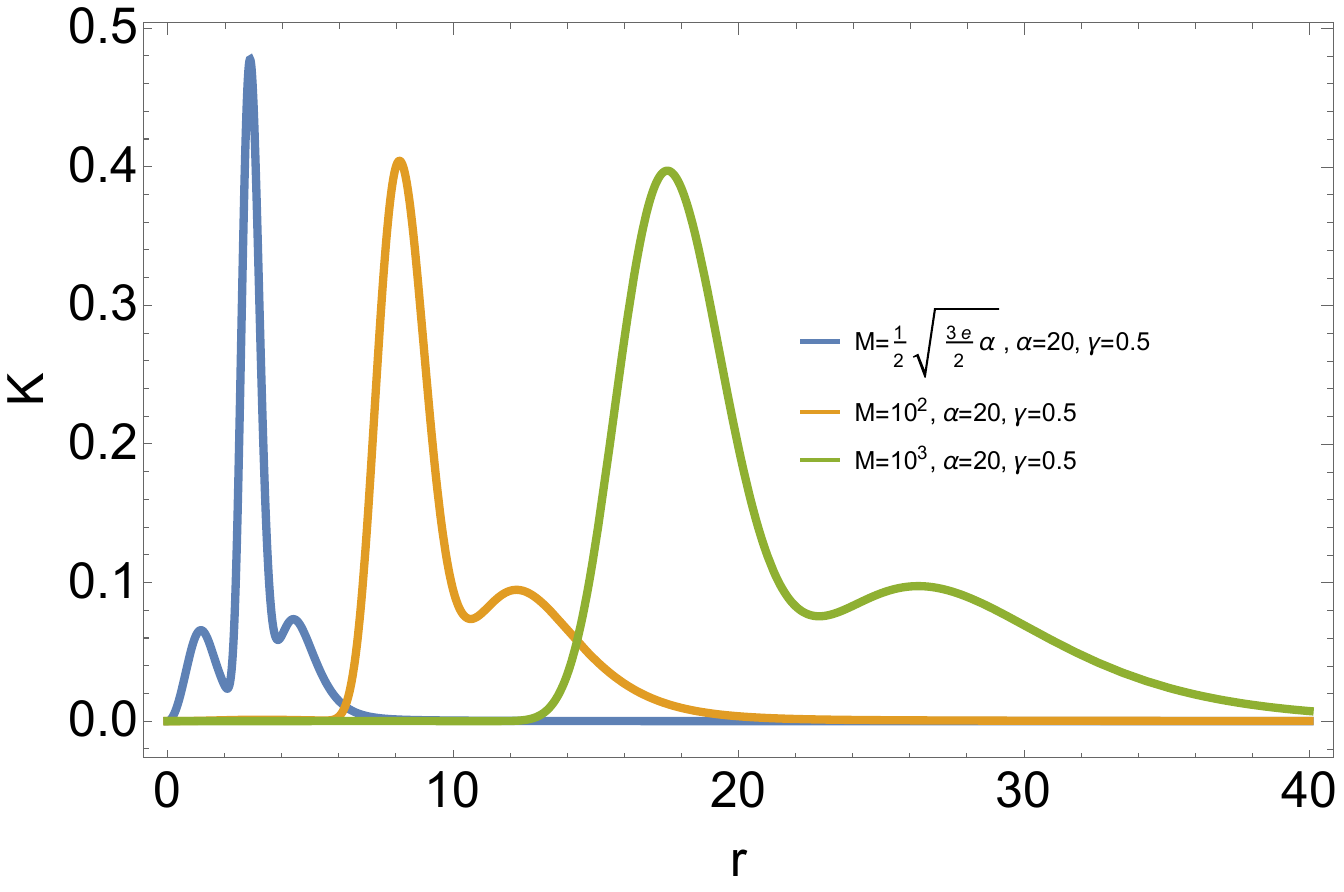}\ \hspace{0.05cm}
  \includegraphics[scale=0.36]{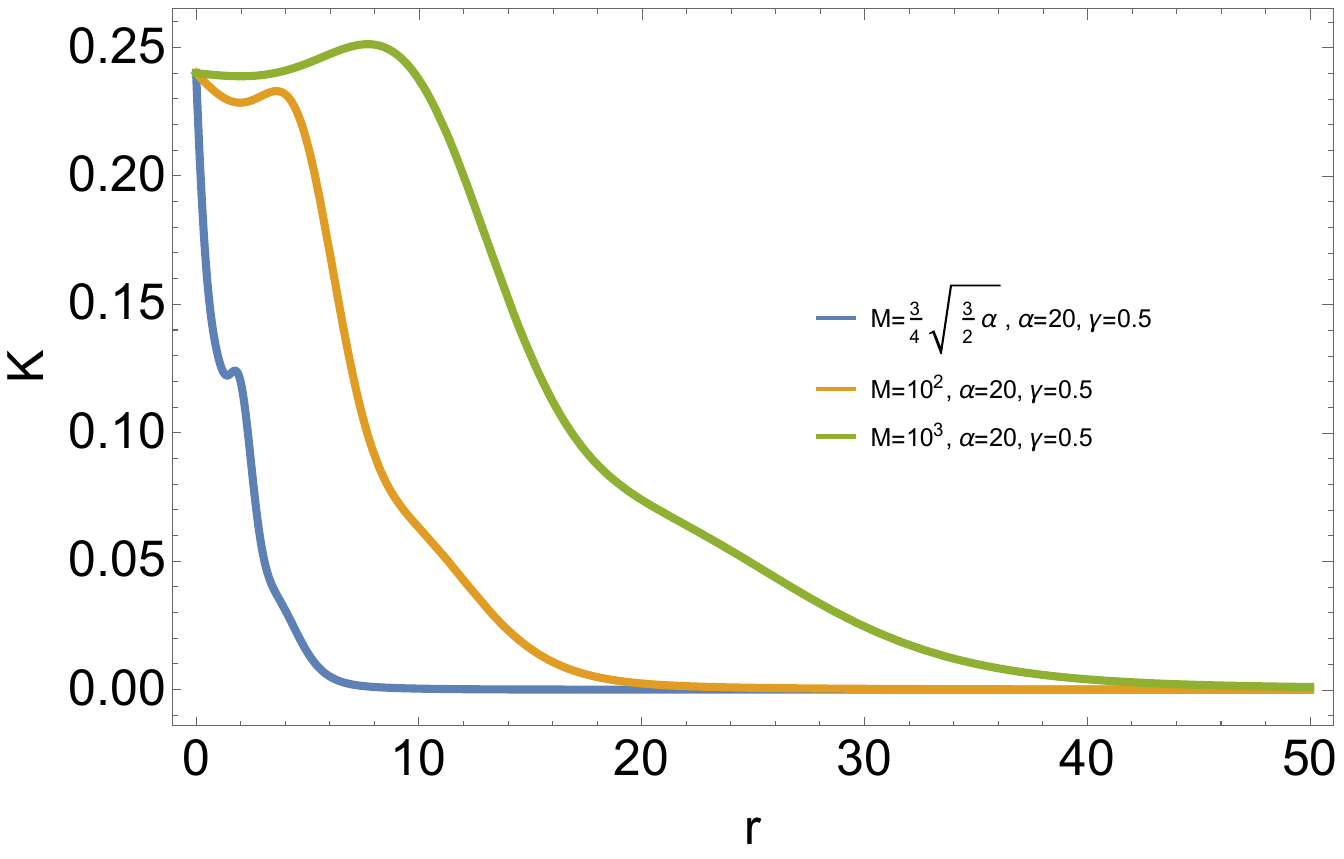}\ \hspace{0.05cm}
\caption{\label{fig7}  The Kretschmann scalar curvature $K$  as
the functions of the radial coordinate $r$.   The left
plot is for the regular black hole with $n=3, \ x=1$, while the
right plot is for Hayward black hole.} }
\end{figure}

We plot  Kretschmann curvature $K$  as the function of the radial
coordinate $r$ in Fig.\ref{fig7}, where $M=\frac{1}{2}
\sqrt{\frac{3 e \alpha}{2}}$ is the minimal value allowed for the
regular black hole with $x=1$ and $n=3$, while $M=\frac{3}{4}
\sqrt{\frac{3 \alpha}{2}}$ is the minimal value allowed for
Hayward black hole. It is found that $K_{max}$ moves to the right
with larger radius as the mass increases for black hole
with $x=1$ and $n=3$, while for Bardeen black hole $K_{max}$ runs
away from the center as well.

\begin{figure} [t]
  \center{
  \includegraphics[scale=0.451]{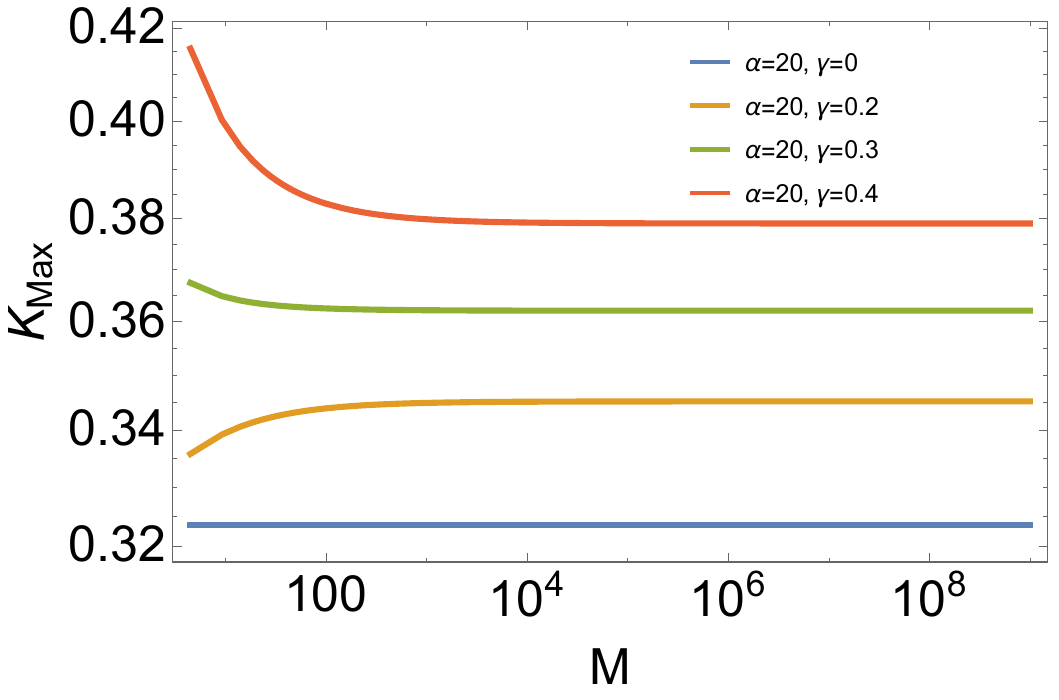}\ \hspace{0.05cm}
  \includegraphics[scale=0.44]{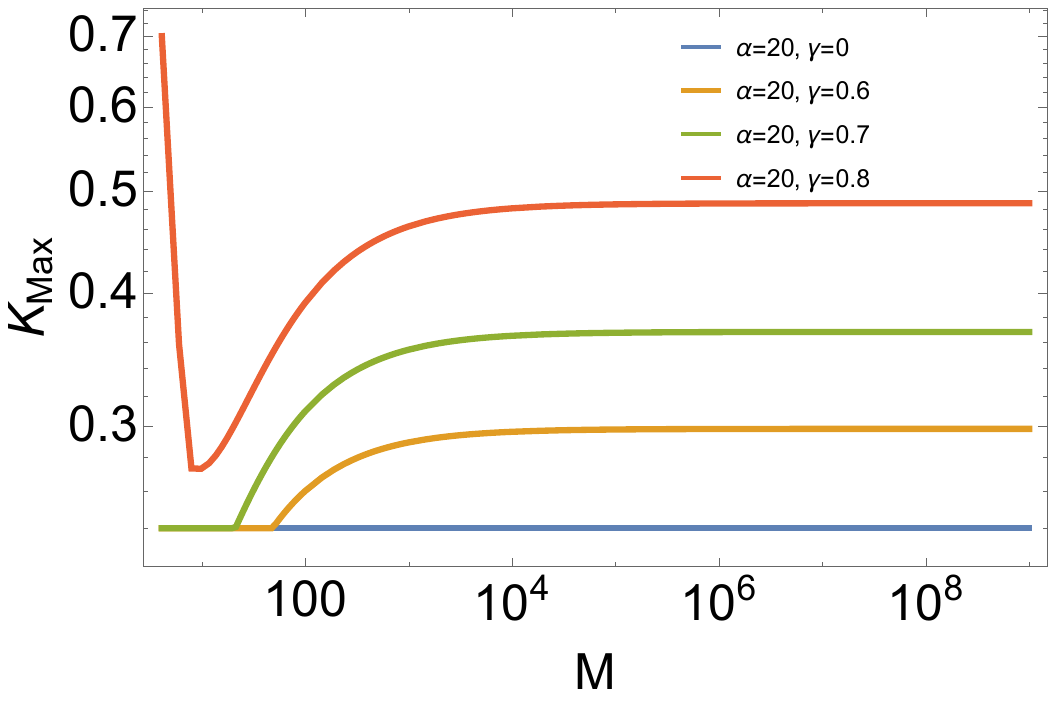}\ \hspace{0.05cm}
\caption{\label{fig7v2} The maximum value of the Kretschmann
scalar curvature $K$ as the function of $M$ for $n=3, \ x=1$(left)
and Hayward black hole(right). }}
\end{figure}

Next we focus on the mass dependent behavior of $K_{max}$. We plot
$K_{max}$ as the function of $M$ for both black holes, as
illustrated in Fig.\ref{fig7v2}. The phenomenon is similar to what
we have found in the previous section. This figure explicitly
exhibits that $K_{max}$ does not depend on the mass of black holes
when $\gamma=0$, as we expect. However, when the time delay
parameter $\gamma$ is turned on, for instance $\gamma=0.2$ in the
left plot
 and $\gamma=0.6, 0.7$
in the right plot, we find $K_{max}$ increases with the mass and
becomes saturated in the large mass limit. Keep increasing
$\gamma$ further, one finds $K_{max}$ becomes larger on the side
with small masses, for instance when $\gamma=0.3,0.4$ in the left
plot and $\gamma=0.8$ in the right plot. In this situation,
$K_{max}$ decreases with the mass and then becomes saturated in
the large mass limit.

Finally, we remark that  the
thermodynamical behavior of this modified regular black hole is quite similar to what we have demonstrated in previous sections.  The maximal value of the Hawking temperature becomes smaller slightly with the increase of $\gamma$, while the entropy becomes larger slightly with the increase of $\gamma$.

\section{Conclusion and Discussion}

In this paper we have developed the regular black holes recently
proposed in \cite{Ling:2021olm} by incorporating the 1-loop
quantum correction to the gravity potential and a time delay
between an observer at the center and the one at infinity. Our
analysis covers two sorts of black holes. One is featured by an
asymptotically Minkowski core, while the other is featured by an
asymptotically de-Sitter core. We have proposed a scheme to figure
out the maximal time delay at the center by scanning the mass of
black holes such that the sub-Planckian feature of Kretschmann
curvature is preserved for black holes during the whole
evaporation process, which greatly improved the strategy presented
in \cite{DeLorenzo:2014pta} where the maximal time delay is
considered for a given mass in the context of modified Hayward
black hole. Therefore, such effective metrics may be more
appropriate to be applied to describe the evolution of Planck
stars or the evaporation of black holes.  We have also compared
the distinct behavior of Kretschmann curvature for these two sorts
of black holes. In general, we have found the mass dependent
behavior of $K_{max}$ becomes complicated when 1-loop quantum
correction and the time delay are taken into account.
Nevertheless, it is still plausible to define a modified regular
black hole with sub-Planckian curvature irrespective of the mass,
thanks to the saturating behavior of $K_{max}$ in the large mass
limit.

Next we remark that the form of $G(r)$ adopted in this paper is
not unique of course. One may consider other forms of $G(r)$, for
instance $G(r)=1-\gamma+\gamma e^{\frac{-\beta M}{\gamma r^3}}$ as
proposed in \cite{DeLorenzo:2015taa}, to introduce 1-loop quantum
correction and the time delay. One would obtain the similar
results. More importantly, we may further improve the form of
effective metric to be more practical to describe the collapse of
star or black holes. One prominent issue in all previous papers is
that one just introduces the time delay by hand, since $\gamma$ is
introduced as a free parameter. A more realistic setup would be
that the time delay is fixed to be a specific one by the
distribution of matter surrounding the center. At the
phenomenological level, it implies that $\gamma$ could be
determined by the other parameters or quantities. For instance, we
could assume that $\gamma=1-(\alpha_{min}/\alpha)^{1/n}$ or
$\gamma=1-(\alpha_{min}/\alpha)^{1/(n-1)}$, then remarkably we
find this value is always smaller than $\gamma_{max}$ for
arbitrary value of $\alpha$ such that a sub-Planckian Kretschmann
scalar curvature is always guaranteed. We expect this conjecture
can be justified  by more detailed investigation in future.

Finally, at the current stage we are strict to consider the regular
black holes with spherical symmetry.  We expect the proposed
construction may be extended to the rotating black holes, as other
authors have performed  for Hayward black holes in
\cite{Bambi:2013ufa,DeLorenzo:2015taa}.

\appendix

\section{The Kretschmann scalar curvature}\label{AppxA}

For ansatz in Eq.(\ref{Eq.metric}),  it is straightforward to derive the Kretschmann
scalar curvature which is given by
\begin{equation}
\begin{aligned}
K &=R^{\mu \nu \rho \lambda} R_{\mu \nu \rho \lambda} \\
&=\frac{(2 \phi (r)+1)^2 G'(r)^4}{4 G(r)^4} +4 \left(\frac{4 \left(r^2 \phi '(r)^2+\phi (r)^2\right)}{r^4}+\phi ''(r)^2\right)    \\
& +\frac{1}{r^2 G(r)^2}(r^2 (2 \phi (r)+1)^2 G''(r)^2+G'(r)^2 (-2 r^2 (2 \phi (r)+1) \phi ''(r) \\
& +9 r^2 \phi
   '(r)^2+8 \phi (r)^2+8 \phi (r)+2)+6 r^2 (2 \phi (r)+1) G'(r) G''(r) \phi '(r)) \\
&-\frac{1}{G(r)^3}((2 \phi (r)+1) G'(r)^2 \left((2 \phi (r)+1) G''(r)+3 G'(r) \phi '(r)\right))\\
&+\frac{1}{r^2 G(r)}(4 \left(r^2 (2 \phi (r)+1) G''(r) \phi
''(r)+G'(r) \phi '(r) \left(3 r^2 \phi ''(r)+4 \phi
   (r)+2\right)\right)).
\end{aligned}
\end{equation}
Furthermore, if we expect that $G(r)$ only plays a role of time
delay at the center, without changing the asymptotical behavior
near the center, it is reasonable to require that
$G'(0)=G''(0)=0$. Then, as $r\rightarrow 0$, we find Kretschmann
scalar curvature behaves as
\begin{equation}
\begin{aligned}
K \sim 4\phi ''(r)^2+ \frac{16 \phi '(r)^2}{r^2} +\frac{16
\phi(r)^2}{r^4}.
\end{aligned}
\end{equation}
Remarkably, we find the function $G(r)$ would not change the value
of Kretschmann scalar curvature at the center, under the condition
that $G'(0)=G''(0)=0$.

\section{The horizon of the modified regular black hole}\label{AppxB}
In this section, we derive the location of the outer horizon for
the modified regular black hole, where the metric components are
specified by Eq.(\ref{Eq.pG}). First of all, we notice that the
location of the outer horizon is not affected by $G(r)$, it is
solely determined by $F(r_h)=0$ and this gives rise to the
relation between $r_h$ and $M$:
\begin{equation}
2M=r_h e^{\alpha M^x / r_h^{n}}.
\end{equation}

We rewrite the radius of the horizon $r_h$ as
\begin{equation}\label{eq_m}
r_h=2 M \left(\frac{\theta}{W(\theta)}\right)^{1/n}, \quad
\theta=-\frac{\alpha}{2 M^{2-x}},
\end{equation}
where
\begin{equation}
W(\theta)=\sum_{n=1}^{\infty} \frac{(-n)^{n-1}}{n !} \theta^{n},
\end{equation}
is the Lambert-W function with $W(\theta)\geq -1$. A real W
requires $\theta \geq -e^{-1}$, thus the mass of modified regular black hole is bounded by
\begin{equation}
M \geq\frac{1}{2} (e \alpha )^{\frac{1}{2-x}}.
\end{equation}

\section{The thermodynamics of the modified black holes}\label{AppxC}
For the modified black holes with non-trivial $G(r)$, the black
hole temperature and the luminosity are respectively given by
\begin{equation}
\begin{aligned}\label{eq_tem}
 T&=\left[-\frac{1}{4 \pi} \sqrt{-g^{tt}g^{rr}} \frac{d}{d r} g_{tt}\right]_{r=r_h}=\frac{\left(2 \phi \left(r_h\right)+1\right) G'\left(r_h\right)+2
   G\left(r_h\right) \phi ' \left(r_h\right)}{4 \pi  \sqrt{G\left(r_h\right)}},\\
L &=\sigma T^{4} A=\frac{\sigma  r_h^2 \left(\left(2 \phi
\left(r_h\right)+1\right) G'\left(r_h\right)+2 G\left(r_h\right)
\phi '\left(r_h\right)\right){}^4}{64 \pi ^3
G\left(r_h\right){}^2}.
\end{aligned}
\end{equation}

Using Eq.(\ref{eq_m}) and (\ref{eq_tem}), one  can derive the heat capacity as
\begin{equation}
C = \frac{d M}{d T} =\frac{d M/d r_h}{d T/ d r_h}.
\end{equation}
Moreover, according to the first law of black hole thermodynamics, the entropy of the modified black hole is given by
\begin{equation}
S=\int \frac{d M}{T}.
\end{equation}
Usually, due to the correction of the Hawking temperature, it is well known that the entropy of modified black holes will deviate from the area law and receive the higher order corrections. For instance, for Hayward black hole in \cite{Li:2016yfd},  the  integrated form of the entropy can be  expanded as
\begin{equation}
S=\pi  r_h^2+\pi  \alpha  \log \left(2 r_h^2-\alpha
   \right)-\frac{\pi  \alpha }{2}-\frac{\pi  \alpha ^2}{2 \left(2 r_h^2-\alpha \right)}.
\end{equation}
For Bardeen black hole in \cite{Sharif:2010pj}, the  integrated form of the entropy  is given by
\begin{equation}
S=2 \pi  \left(\left(\frac{r_h}{2}-\frac{\alpha }{r_h}\right) \sqrt{\alpha +r_h^2}+\frac{3}{2} \alpha  \log
   \left(\sqrt{\alpha +r_h^2}+r_h\right)\right).
\end{equation}
For regular black hole  in \cite{Xiang:2013sza}, the  integrated form of the entropy  can be  written  as
\begin{equation}
S=\int \frac{d M}{T}=e^{\frac{\alpha }{r_h^2}} \pi  r_h^2+2 \pi  \alpha  \int \frac{e^{\frac{\alpha }{r_h^2}}}{r_h} \, dr_h.
\end{equation}
For the black hole presented in this paper, such expansions become very complicated and we just present the numerical results in the main body of the text.

\centerline{\rule{80mm}{0.5pt}}

\end{document}